\documentclass[
amsmath,amssymb,pra,
 twocolumn
 ]{revtex4}
\usepackage{graphicx}% Include figure files
\usepackage{dcolumn}% Align table columns on decimal point
\usepackage{bm}% bold math
\usepackage{hyperref}% add hypertext capabilities
\usepackage{bbold}

\renewcommand{\d}[2]{\ensuremath{\frac{\text{d} #1}{\text{d} #2}}}

\newcommand{\ket}[1]{\ensuremath{\left| #1 \right>}}

\renewcommand{\exp}[1]{\ensuremath{ \; \text{exp} \left( #1 \right) } }

\begin{document}

\preprint{APS/123-QED}

\title{Quantum impurity in a Luttinger liquid  -\\  \quad \quad Exact Solution of the Kane-Fisher Model}% Force line breaks with \\
\author{Colin Rylands} 
\email{rylands@physics.rutgers.edu}
\author{Natan Andrei}
\email{natan@physics.rutgers.edu}
\affiliation{Department of Physics, Rutgers University,
Piscataway, New Jersey 08854.
}

\date{\today}% It is always \today, today,
%             %  but any date may be explicitly specified
%
\begin{abstract}
A Luttinger Liquid coupled to a quantum  impurity describes a large number of physical systems.  The Hamiltonian consists of  left-  and  right-moving  fermions  
 interacting  among themselves via a density-density coupling and scattering off a localised transmitting and reflecting impurity.  We solve exactly the Hamiltonian by means of an incoming-outgoing  scattering Bethe basis  which properly incorporates all scattering processes.  A related model, the Weak-Tunnelling model, wherein the impurity is replaced by a tunnel junction, is solved by the same method.   The  consistency of the construction is established 
 through  a generalised Yang-Baxter relation. Periodic boundary conditions are imposed and the resulting Bethe Ansatz equations are derived  by means of the Off Diagonal Bethe Ansatz approach.  We derive the spectrum of the model for all coupling constant regimes  and calculate the impurity free energy.  We discuss  the low energy behaviour of the systems for both  repulsive and attractive interactions.
\end{abstract}

\maketitle

\section{Introduction}
It has long been known that interactions can have drastic effects in low dimensional systems \cite{TG}. A striking example of this was elucidated by Kane and Fisher \cite{KF}. It was shown that a local impurity can be a relevant or irrelevant perturbation to a Luttinger Liquid depending on the sign of the interaction in the liquid. For repulsive  interactions  amongst the fermions  the strength of the impurity will grow at low energy  and the one dimensional system will  be split into two Luttinger liquids  weakly coupled at their edges by a tunnelling term (Weak-Tunnelling Hamiltonian), while for  attractive interactions the strength of the impurity will decrease and the system  will heal itself. Hence one finds   a vanishing conductance at the impurity site at low temperature in the first case and in a perfect conductance in the second. 

This has implications for many experimentally realisable  quantum systems. Amongst these are chiral edge states of Quantum Hall materials \cite{LLRMP} and electronic quantum circuits\cite{circuit}.   More exciting perhaps is the possibility to realise such a system with cold atomic gases \cite{IBRMP}. The measure of control afforded by these experiments in addition to the ability to tune parameters including the interaction strength makes this the perfect setting to study the effects of interactions on a localised impurity. Isolated one dimensional systems are readily achievable and recent advances have made it possible to study transport albeit with 2 dimensional leads \cite{Coldatom} \cite{esslinger}.
In such isolated quantum systems, integrability also has a large effect. The existence of a large number of conserved quantities strongly constrains the dynamics \cite{NC} and will have implications for transport. 

In this article we introduce a new type of coordinate Bethe Ansatz for use in quantum impurity models with bulk interaction. We present the method by solving exactly the Kane-Fisher model of an impurity in a Luttinger liquid with arbitrary boundary conditions. The method uses a scattering Bethe basis which incorporates the impurity scattering processes that lead to a varying number of left and right movers. The boundary condition problem leads to a Quantum Inverse Scattering problem which is in turn solved using the Off Diagonal Bethe Ansatz (ODBA) \cite{ODBA} approach of deriving the Bethe Ansatz equations.  It has the advantage that it  does not require an explicit reference state and so is suited to problems where it is absent, which is the case in the present model.  Incorporating twisted boundary conditions being physically equivalent to driving a persistent current around the system allows for the possibility of studying transport across the impurity.  

We also study the Weak-Tunnelling Hamiltonian describing two separate Luttinger liquids coupled via a tunnelling parameter. The model is of great interest by itself and  is thought to describe the strong coupling fixed point of the Kane-Fisher model.  We find that the Weak-Tunnelling Hamiltonian is solvable by the same procedure requiring only simple modifications and  show it is dual to the impurity model. 

The rest of the article is organised as follows: In section II  we introduce the scattering Bethe basis  which incorporates the impurity's selecting - scattering mechanism  and prove it's  consistency by introducing a generalization  of  the Yang-Baxter and reflection equations.  In section III we provide a similar construction for the Weak-Tunnelling Hamiltonian.
The spectrum of the model is found in section IV. The system of Bethe Ansatz equations  is shown to be formally similar to that of the open XXZ model with  boundary terms.  One diagonal boundary corresponds to the twist and the other describes the impurity. Using the ODBA we are able to obtain the eigenvalues and Bethe equations. The thermodynamics of the model are discussed in section V  where  we calculate the free energy and specific heat of the impurity as well as the difference in the impurity entropy in the UV and IR when interactions are repulsive.  The Weak-Tunnelling Hamiltonian is examined, its complementarity with the Kane-Fisher model is shown and the thermodynamics in the attractive regime briefly discussed.

\section{Bethe Basis of the impurity-Luttinger model}
The Hamiltonian of the impurity model we seek to diagonalise is $H=H_{k}+H_{g}+H_I$ with the various terms given by,
\begin{eqnarray}\label{H}
H_{k}&=&\sum_{\sigma=\pm,a=\uparrow,\downarrow}\int \sigma\psi^\dag_{\sigma,a}\left(-i\partial_x-  \cal{A} \right)\psi_{\sigma,a}(x),\\
H_g&=&\sum_{a,b}4g\int \psi_{+,a}^\dag\psi^\dag_{-,b}\psi_{-,b}\psi_{+,a},\\\nonumber
H_I&=&\sum_aU \left[\psi^\dag_{+,a}(0)\psi_{-,a}(0)+\psi_{-,a}^\dag (0)\psi_{+,a} (0)\right]\\
&&+U' \left[\psi^\dag_{+,a}(0)\psi_{+,a}(0)+\psi_{-,a}^\dag (0)\psi_{-,a} (0)\right].
\end{eqnarray}
Here $\psi^\dag_{\pm,a},~\psi_{\pm,a}$ with $a=\uparrow,\downarrow$ are creation operators for the right ($+$) and left ($-$) moving  fermions with spin, $U'$ and $U$ describe the forward and backward scattering off the impurity respectively and $g$ is the fermion-fermion interaction strength. We have set $v_f=1$ and $\epsilon_f=0$. In addition have included a gauge field $\mathcal{A}$ which, when the system is placed on a ring means it is threaded by a flux $\Phi=\int_x  \mathcal{A}$. Equivalently we may  solve for the wavefunction with  twisted boundary conditions. This will induce a persistent current throughout the system and allow the effect of the impurity on the current to be studied. Since we have chosen the interaction to be isotropic in spin we will assume these indices as implicit in what follows. 
 
 To begin we discuss the construction of the eigenfunctions of $H$.  In the presence of the impurity only the total number of fermions $N=N_++N_-$ is conserved, hence the wave functions must
 consist of components of left and right movers consistent with $N$. We start with the single particle eigenstates, the most general form for which can be written as
\begin{eqnarray}\nonumber
\int \mathrm{d}x \left[\left(e^{ikx}A^{[10]}_+\psi^\dag_+(x)+e^{-ikx}A^{[10]}_-\psi^\dag_-(x)\right)\theta(-x)\right.\\\label{n1}
\left.+\left(e^{ikx}A^{[01]}_+\psi^\dag_+(x)+e^{-ikx}A^{[01]}_-\psi^\dag_-(x)\right)\theta(x)\right]\ket{0}.
\end{eqnarray}
Applying the Hamiltonian to the wave function fixes two of these amplitudes $A_\pm^{[\cdot\cdot]}$. Here we wish to take a physical picture and define a $S^{10}$ which maps  a particle past the impurity. This is in contrast to what is standard in Bethe ansatz where the S-matrix maps between regions of configuration space to the left and right of the impurity. Therefore we consider $A_+^{[10]}$and $A_-^{[01]}$ as the incoming amplitudes and  $A_-^{[10]}$and $A_+^{[01]}$ as the outgoing ones. The solution of the Schrodinger equation relates the two sets via
\begin{eqnarray}\label{s}
\begin{pmatrix}
A^{[01]}_+\\
A^{[10]}_-
\end{pmatrix}=S
\begin{pmatrix}
A^{[10]}_+\\
A^{[01]}_-
\end{pmatrix},~~
S=\begin{pmatrix}
\alpha && \beta\\
\beta &&\alpha
\end{pmatrix},\\~~~~\alpha=\frac{1-U^2/4+U'^2/4}{1+i U'+U^2/4-U'^2/4},\\
\beta=\frac{-iU}{1+i U'+U^2/4-U'^2/4}.
\end{eqnarray}
We recognise $\alpha$ and $\beta$ as the transmission and reflection coefficients respectively and note the unimportant role of the forward scattering term. Its presence merely redefines these coefficients but does not change the left-right mixing imposed by the backward scattering term. In what follows we set $U'=0$.

  The form in which we have written the above equation allows us to easily apply periodic or twisted boundary conditions,
\begin{equation}\label{PBc}
e^{-ikL}\begin{pmatrix}
A^{[10]}_+\\
A^{[01]}_-
\end{pmatrix}=\begin{pmatrix}
e^{i\Phi} && 0\\
0 &&e^{-i\Phi}
\end{pmatrix}
S
\begin{pmatrix}
A^{[10]}_+\\
A^{[01]}_-
\end{pmatrix}.
\end{equation}

We now proceed to the two particle case.  The interaction term $ H_g$  couples left- to right-movers only  and preserves their number unchanged unlike the impurity term.  Thus in the absence of the impurity  a state consisting of one left mover and one right mover takes the form $|F^{L,R}\rangle = \int dx \, dy\, F(x,y)
 \psi^{\dagger}_+(x) \psi^{\dagger}_-(y) |0\rangle$, where the wave function $F(x,y)$ must satisfy the eigenvalue equation, 
 \begin{eqnarray}\nonumber
[-i(\partial_x - \partial_y)  + 4 g \delta(x-y)] F(x,y)= EF(x,y)
\end{eqnarray}
The solution is easily found to be
\begin{eqnarray}\nonumber
F(x,y)= A e^{ik_1x-ik_2 y}[\theta(x-y) +e^{i\phi} \theta(y-x)]
\end{eqnarray}
and the scattering phase shift  given by 
 \begin{eqnarray}\nonumber
 e^{i\phi}=\frac{1-ig}{1+ig}.
 \end{eqnarray}
For the scattering of two right movers or two left movers the phase shift is actually undetermined by the Schrodinger equation, we choose it to be: $ e^{i\phi_{++} }=   e^{i\phi_{--}} =1$.
 
  As seen for a single particle the impurity mixes both the left and right movers. A non-interacting model could therefore  be handled via utilising an odd-even basis    $\psi_{e/o}(x)=(\psi_+(x) \pm \psi_-(-x))/\sqrt2$. However doing so for the full model will only serve to complicate the interaction term. On the other hand in the absence of the impurity the left-right basis is appropriate. To diagonalise both we need to use a basis which naturally incorporates both aspects,  we'll refer to it as an in-out  scattering Bethe Basis. 
  
  To construct it we divide configuration space into 8 regions, to be labelled $Q$ , which are specified not only by the ordering of $x_1$, $x_2$ and  the impurity but also according to which position is closer to the origin. For example if $x_1$ is to the left of the impurity, $x_2$ to its right  with $x_2$ closer  to the impurity then the amplitude in this region is denoted $A^{[102B]}_{\sigma_1\sigma_2}$, $\sigma_j=\pm$ being the chirality of the particle at $x_j$. The region in which $x_1$ is closer is denoted  $A^{[102A]}_{\sigma_1\sigma_2}$.  The consequence for the wavefunction is that we include Heaviside functions $\theta (x_Q)$ which have support only in a certain region, e.g $\theta(x_{[102B]})=\theta(x_2)\theta(-x_1)\theta (-x_1-x_2)$. A general two particle eigenstate for $H$ can be written as,
\begin{eqnarray}\nonumber
\ket{k_1,k_2}=\sum_Q\sum_{\sigma_1\sigma_2}\int\theta (x_Q)A_{\sigma_1\sigma_2}^{Q}e^{\sigma_1ik_1x_1+\sigma_2ik_2x_2}\\
\times\psi^\dag_{\sigma_1}(x_1)\psi^\dag_{\sigma_1}(x_2)\ket{0}.
\end{eqnarray}
The form of this wavefunction requires some comment. The linear derivative acts as $\pm i(\partial_1-\partial_2)$ when the particles are of opposite chirality and as $\pm i(\partial_1+\partial_2)$ when they have the same chirality. This allows us to introduce an arbitrary function of $x_1\pm x_2$ when the particles are of the same or opposite chirality. Accordingly, applying the Hamiltonian to this ansatz fixes some but not all the amplitudes.  In particular when switching between the regions weighted by $\theta(\pm(x_1-x_2))$ in the $\sigma_1=\sigma_2$ sector and   $\theta(\pm(x_1+x_2))$ in the  $\sigma_1=-\sigma_2$ sector the linear derivative allows us to choose any S-matrix we like provided it does not mix the   $\sigma_1=\sigma_2$ with the  $\sigma_1=-\sigma_2$ amplitudes \footnote{This procedure is actually very natural and  is required whenever  a degenerate level is perturbed. In our case, the energy level $k_1+k_2$  is  degenerate with $(k_1+q)+(k_2-q)$ for any $q$. Thus, as degenerate perturbation theory requires,  an appropriate basis in the degenerate subspace needs to be found in which the perturbation can be turned on. This corresponds to the consistent choice of the S-matrices, as described}.  The specific form of this additional S-matrix is dictated by the requirement that the wavefunction be consistent. Typically this would require the S-matrices be solutions of the Yang Baxter equation but here the different configuration space set up will modify this and will lead to a generalised Yang-Baxter relation. To make these statements more explicit let us form column vectors of the amplitudes,
\begin{eqnarray}
\centering\nonumber
&&\vec{A}_1=\begin{pmatrix}
A_{++}^{[120B]}\\
A_{+-}^{[102B]}\\
A_{-+}^{[201B]}\\
A_{--}^{[021B]}
\end{pmatrix}~\vec{A}_2=\begin{pmatrix}
A_{++}^{[210A]}\\
A_{+-}^{[102A]}\\
A_{-+}^{[201A]}\\
A_{--}^{[012A]}
\end{pmatrix}~\vec{A}_3=\begin{pmatrix}
A_{++}^{[201A]}\\
A_{+-}^{[012A]}\\
A_{-+}^{[210A]}\\
A_{--}^{[102A]}
\end{pmatrix}\\\nonumber
&&\vec{A}_4=\begin{pmatrix}
A_{++}^{[201B]}\\
A_{+-}^{[021B]}\\
A_{-+}^{[120B]}\\
A_{--}^{[102B]}
\end{pmatrix}
~\vec{A}_5=\begin{pmatrix}
A_{++}^{[021B]}\\
A_{+-}^{[201B]}\\
A_{-+}^{[102B]}\\
A_{--}^{[120B]}
\end{pmatrix}
~\vec{A}_6=\begin{pmatrix}
A_{++}^{[012A]}\\
A_{+-}^{[201A]}\\
A_{-+}^{[102A]}\\
A_{--}^{[210A]}
\end{pmatrix}
\\&&~~~~~~\vec{A}_7=\begin{pmatrix}
A_{++}^{[102A]}\\
A_{+-}^{[210A]}\\
A_{-+}^{[012A]}\\
A_{--}^{[201A]}
\end{pmatrix}
~~~~~~~\vec{A}_8=\begin{pmatrix}
A_{++}^{[102B]}\\
A_{+-}^{[120B]}\\
A_{-+}^{[021B]}\\
A_{--}^{[201B]}
\end{pmatrix}
\end{eqnarray} 
We interpret $\vec{A}_1$ ($\vec{A}_2$) as the amplitudes where both particles are incident on the impurity but particle 2 (1) is closer,  $\vec{A}_5$ ($\vec{A}_6$) are the amplitudes in which both particles are outgoing with particle 2 (1) closer to the impurity, $\vec{A}_8$ ($\vec{A}_3$) describes  particle 2 (1)  having scattered off the impurity and is still closer to the impurity than 1 (2) while $\vec{A}_7$  ($\vec{A}_4$ )  also describes particle 2 (1)  having scattered but with 1 (2) is closer. The Hamiltonian fixes the following relations between these amplitudes
\begin{eqnarray}
\vec{A}_8=S^{20}\vec{A}_1,~~~\vec{A}_3=S^{10}\vec{A}_2,\\
\vec{A}_5=S^{20}\vec{A}_4,~~~\vec{A}_6=S^{10}\vec{A}_7,\\
\vec{A}_7=S^{12}\vec{A}_8,~~~\vec{A}_4=S^{12}\vec{A}_3,\\\label{S1}
\end{eqnarray}
where 
\begin{eqnarray}
S^{20}=S\otimes \mathbb{1},~~~S^{10}=\mathbb{1}\otimes S,\label{S}
\end{eqnarray}
and, as discussed above,
\begin{eqnarray}
S^{12}=\begin{pmatrix}
1&&0&&0&&0\\
0&&e^{i\phi}&&0&&0\\
0&&0&&e^{i\phi}&&0\\
0&&0&&0&&1
\end{pmatrix}.
\end{eqnarray}
The freedom mentioned previously enters upon considering $ \vec{A}_1\leftrightarrow \vec{A}_2$ and $ \vec{A}_5\leftrightarrow \vec{A}_6$. Again, these S-matrices are  restricted only in that they cannot mix $\sigma_1=\sigma_2$ amplitudes with $\sigma_1=-\sigma_2$.
We choose to take
\begin{eqnarray}
\vec{A}_2=W^{12}\vec{A}_1,~~~\vec{A}_6=W^{12}\vec{A}_5,\\\label{W}
W^{12}=\begin{pmatrix}
1&&0&&0&&0\\
0&&0&&1&&0\\
0&&1&&0&&0\\
0&&0&&0&&1
\end{pmatrix}.
\end{eqnarray} This is dictated  by the consistency of the wave function which requires the S-matrices to satisfy a reflection equation,
\begin{equation}
S^{20}S^{12}S^{10}W^{12}=W^{12}S^{10}S^{12}S^{20}
\end{equation}
\begin{figure}
\includegraphics[trim=30 30 30 30,width=.5\textwidth]{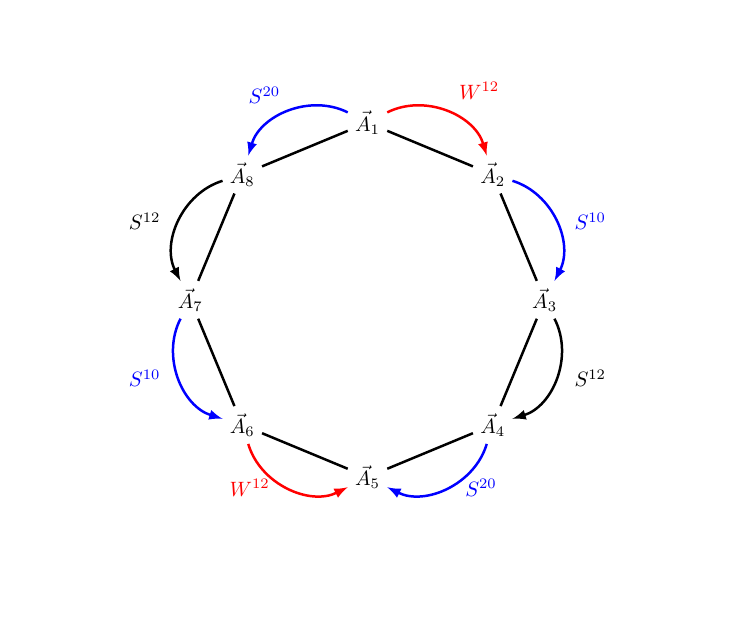}\label{fRE}
\caption{(Color Online)The amplitudes are related by applying the operators as depicted here. For consistency we require the amplitudes obtained by proceeding clockwise or counter-clockwise are the same resulting in \eqref{RE}.}
\end{figure}
Inserting \eqref{W}\eqref{S}\eqref{S1} it is easy to see this indeed holds. A schematic representation  is given in the Fig 1.  By introducing the extra regions indexed by $A,B$ we have changed the consistency condition from the Yang-Baxter equation to a generalised version  that takes the form of a  reflection equation. As explained, the partition to these extra regions is dictated by  linear derivative and the degeneracies associated with it,  which require  us to choose  the correct basis in the degenerate subspace. This basis, the Bethe basis, corresponds to the introduction of the  S-Matrix  $W^{12}$ which satisfies the consistency conditions. Such a  degeneracy is not present in a massive theory  in which case integrability  is inconsistent  with a  nontrivial  bulk  interaction in the presence of a transmitting and reflecting impurity  \cite{mussardo}.

The generalisation to $N$ particles is immediate. The $N$ particle eigenstate with energy $E=\sum_j^N k_j$ is,
\begin{eqnarray}
\ket{\vec{k}}=\sum_Q\sum_{\vec{\sigma}}\int\theta (x_Q)A_{\vec{\sigma}}^{Q}e^{i\sum \sigma_j k_jx_j}\prod \psi^\dag_{\sigma_j}(x_j)\ket{0}.
\end{eqnarray}
The sum is over the $2^N N!$ regions consisting of all  orderings of $x_j$ and the origin and indexed by which particle is closest to the impurity. Just as in the two particle case the amplitudes $A_{\vec{\sigma}}^{Q}$ are related to each other by applying the S-matrices,
\begin{eqnarray}
S^{j0}&=&S_j\otimes_{k\neq j}\mathbb{1},\\\label{Sij}
S^{ij}&=&\begin{pmatrix}
1&&0&&0&&0\\
0&&e^{i\phi}&&0&&0\\
0&&0&&e^{i\phi}&&0\\
0&&0&&0&&1
\end{pmatrix}_{ij}\otimes_{k\neq i,j}\mathbb{1},~~\\\label{Wij}
~~W^{ij}&=&\begin{pmatrix}
1&&0&&0&&0\\
0&&0&&1&&0\\
0&&1&&0&&0\\
0&&0&&0&&1
\end{pmatrix}_{ij}\otimes_{k\neq i,j}\mathbb{1}.
\end{eqnarray}
The subscripts denote which particle spaces the operators act upon. In order for this wavefunction to be consistent it must satisfy the following Yang-Baxter and reflection equations,
\begin{eqnarray}\label{RE}
S^{k0}S^{jk}S^{j0}W^{jk}&=&W^{jk}S^{j0}S^{jk}S^{k0}\\\label{YB1}
W^{jk}W^{jl}W^{kl}&=&W^{kl}W^{jl}W^{jk}\\\label{YB2}
W^{jk}S^{jl}S^{kl}&=&S^{kl}S^{jl}W^{jk}.
\end{eqnarray} 
 Satisfying these is a sufficient condition for the consistency of the wave function because the S-matrices form a representation of the reflection group just as those in other integrable models form a representation of the permutation group \citep{ZinnJustin}.  This will be made evident in the next section when the continuous versions of the S-matrices and the Bethe equations are found.

To determine the thermodynamic spectrum of the model we place the system on a ring of size $L$. The flux, $\Phi=\mathcal{A} L$ through the loop then imposes twisted boundary conditions so that upon traversing the entire system a particle picks up an additional phase $e^{\sigma i\Phi}$, $\sigma$ being the chirality of the particle.  We obtain the following equations which determine $k_j$
\begin{eqnarray}\label{PBC}
&& e^{-ik_jL}A_{\sigma_1\dots\sigma_N}=\left(Z_j\right)^{\sigma'_1\dots\sigma'_N}_{\sigma_1\dots\sigma_N}A_{\sigma'_1\dots\sigma'_N}\\
 &&Z_j=W^{j-1j}.. W^{1j}B_jS^{1j}.. S^{jN}S^{j0}W^{jN}.. W^{jj+1}
\end{eqnarray} 
where the matrix $Z_j$ transfers the $j$th particle around the ring. Here the matrices $B_j$ act in the  $j$th particle chirality space and impose the twisted boundary conditions,
\begin{equation}
B_j=\begin{pmatrix}
e^{i\Phi}&&0\\0&&e^{-i\Phi}
\end{pmatrix}.
\end{equation}
Alternatively we could require hard wall boundary conditions at $x=\pm L/2$ by taking $B_j=\sigma_x$.

 Using \eqref{RE}\eqref{YB1}\eqref{YB2} it can be shown  that all transfer matrices $Z_j$ are equivalent and so we restrict our attention to solving,
\begin{eqnarray}\nonumber
 \left(B_1S^{12}\dots S^{1N} S^{10}W^{1N}\dots W^{12}\right)^{\sigma'_1\dots\sigma'_N}_{\sigma_1\dots\sigma_N}A_{\sigma'_1\dots\sigma'_N}\\\label{Z}
 = e^{-ikL}A_{\sigma_1\dots\sigma_N}.
\end{eqnarray}
This is a feature of many quantum impurity models. It arises due to the lack of a dimensionful parameter in the Hamiltonian which results in S-matrices which are $k$ independent. We denote the operator on the left hand side $Z$.  Its eigenvalues determine the allowed values of the momenta $k_j$ and therefore  the spectrum, $E= \sum_j k_j$. However, before proceeding to the diagonalization of the transfer matrix we turn to the solution  of another closely related model, the Weak-Tunnelling model.

\section{Bethe Ansatz eigenstates of the Weak -Tunnelling Hamiltonian} 
The embedding of an impurity in a Luttinger liquid could be viewed from the complementary scenario of two liquids which are coupled by a weak link or tunnel junction.
Therefore in addition to the impurity model we will also consider the Weak-Tunnelling Hamiltonian, $H_{WT}$ which is believed to govern the behaviour of the system in the vicinity of the strong coupling point. It includes two Luttinger liquids each described by $H_k+H_g$, occupying the regions from $-L/2$ to $0$ and $0$ to $L/2$ denoted by the subscripts $l$ and $r$ respectively. These are coupled to each other via the tunnelling term,
\begin{eqnarray}
H_t=t\big(\psi^\dag_{+,r}(0)+\psi^\dag_{-,r}(0)\big)\big(\psi_{+,l}(0)+\psi_{-,l}(0)\big)+\text{h.c} \;\;\;
\end{eqnarray}
which allows for tunnelling between the otherwise disjoint Luttinger liquids. 

The single particle solution of the Weak-Tunnelling Hamiltonian is of a similar form to \eqref{n1},
\begin{eqnarray}\nonumber
&&\int_{-\frac{L}{2}}^0 \left[e^{ikx}A^{[10]}_+\psi^\dag_{+,l}(x)+e^{-ikx}A^{[10]}_-\psi^\dag_{-,l}(x)\right]\ket{0}\\
&&+\int^{\frac{L}{2}}_0 \left[e^{ikx}A^{[01]}_+\psi^\dag_{+,r}(x)+e^{-ikx}A^{[01]}_-\psi^\dag_{-,r}(x)\right]\ket{0}.\;\;\;\;
\end{eqnarray} 
 Here we have used the same notation as in the impurity case so that $A_{\sigma}^{[10]}$ is the amplitude of a particle of chirality $\sigma$ in the left system and  $A_{\sigma}^{[01]}$ in the right system. Acting on this with the Hamiltonian  and using the boundary conditions $\psi^\dag_{+,l}(0)=\psi^\dag_{-,l}(0)$ and $\psi^\dag_{+,r}(0)=\psi^\dag_{-,r}(0)$ we find that
\begin{eqnarray}\label{t}
\begin{pmatrix}
A^{[01]}_+\\
A^{[10]}_-
\end{pmatrix}=S_t
\begin{pmatrix}
A^{[10]}_+\\
A^{[01]}_-
\end{pmatrix}
&&,~~S_t=\begin{pmatrix}
\alpha_t && \beta_t\\
\beta_t &&\alpha_t
\end{pmatrix},\\~~~~\alpha_t=\frac{-4it}{1+4t^2},~&&~~~\beta_t=\frac{1-4t^2}{1+4t^2}.
\end{eqnarray}
The imposition of hard wall boundary conditions at $x=\pm L/2$ gives this time
\begin{eqnarray}
e^{-ikL}\begin{pmatrix}
A^{[01]}_+\\
A^{[10]}_-
\end{pmatrix}=\sigma_x
S_t
\begin{pmatrix}
A^{[01]}_+\\
A^{[10]}_-
\end{pmatrix}.
\end{eqnarray} 

The set up for higher particle number is the same as for the impurity model and the analysis of the preceding section transfers to the present case. This enables us to construct consistent $N$ particle eigenstates. The two particle S-matrices are given by \eqref{Sij} and \eqref{Wij}. The difference is the single particle S-matrix  $S^{j0}$ being replaced with $S^{j0}_t=S_{t\,j}\otimes^N_{k\neq j}\mathbb{1}$. These are readily seen to satisfy the consistency conditions \eqref{RE}-\eqref{YB2}. 

 As before we impose boundary conditions to determine the spectrum and obtain for hard walls at $x=\pm L/2$,
\begin{eqnarray}\nonumber
 \left(\sigma_x S^{12}\dots S^{1N} S_t^{10}W^{1N}\dots W^{12}\right)^{\sigma'_1\dots\sigma'_N}_{\sigma_1\dots\sigma_N}A_{\sigma'_1\dots\sigma'_N}\\\label{Zt}
 = e^{-ikL}A_{\sigma_1\dots\sigma_N}.
\end{eqnarray}
We could also have applied periodic or twisted boundary conditions by including $B_1$ instead of $\sigma_x$. The system with periodic or twisted boundary conditions no longer describes two disjoint liquids filling the left and right half lines but rather a ring containing a weak link. This is the dual system to the impurity model on a ring. To distinguish  with the impurity model we denote the operator  above by $Z_t$.

In what follows we will be concerned with properties of the impurity and weak link which will be independent of the type boundary condition imposed. 

\section{Off-Diagonal Bethe Ansatz}
In the previous section we showed that in order to determine the spectrum of $H$ or $H_{WT}$ we must diagonalise  $Z$  or $Z_t$. To achieve this we will make use of the Off Diagonal Bethe Ansatz \cite{ODBA}. This method allows one to determine the eigenvalues and eigenvectors of a transfer matrix when a proper reference state is absent. It has already been successfully used to obtain the exact solutions for many integrable models with a broken $U(1)$ symmetry. The present problem will be shown to be mappable onto one arising when an XXZ  Hamiltonian is diagonalised with open boundary conditions, which is amongst those already considered\cite{ODBAXXZ}. We will use its solution to obtain the eigenvalues of $Z$ and $Z_t$. Although the following procedure can be used with any type of boundary conditions we will do so only for twisted boundary conditions.   

We begin by constructing the monodromy matrix, the central object of the  quantum inverse scattering (QIS) and of the ODBA approaches. It  is formed from an XXZ  - like R-matrix
and of reflection matrices.  The R-matrix is
\begin{equation}\label{R}
\mathcal{R}(u)=\begin{pmatrix}
1&&0&&0&&0\\
0&&\frac{\sinh{u}}{\sinh{(u+\eta)}}&&\frac{\sinh{\eta}}{\sinh{(u+\eta)}}&&0\\
0&&\frac{\sinh{\eta}}{\sinh{(u+\eta)}}&&\frac{\sinh{u}}{\sinh{(u+\eta)}}&&0\\
0&&0&&0&&1
\end{pmatrix}.
\end{equation}
where $u$ is the spectral parameter and $\eta$ the crossing parameter which encodes the interactions of the model. We shall identify it in our case as : $e^{-\eta}=e^{i\phi}= \frac{1-ig}{1+ig}$ with $g$ the Luttinger liquid interaction coupling constant.

The reflection or boundary matrices, $K^\pm(u)$, we use take the form of integrable boundary conditions for the XXZ model \cite{DeVega} with components,
\begin{eqnarray}\label{K}
K^-_{11}(u)&=&K^-_{22}(u)=2i\cosh{(c+\theta/2)}\cosh{u}\\
K^-_{12}(u)&=&K^-_{21}(u)=\sinh{2u}\\\nonumber
K^+_{11}(u)&=&2\left(\sinh{(-\theta)}\cosh{(i\Phi)}\cosh{(u+\eta)}\right.\\&&\left.+\cosh{(\theta)}\sinh{(i\Phi)}\sinh{(u+\eta)}\right)\\\nonumber
K^+_{22}(u)&=&2\left(\sinh{(-\theta)}\cosh{(i\Phi)}\cosh{(u+\eta)}\right.\\&&\left.-\cosh{(\theta)}\sinh{(i\Phi)}\sinh{(u+\eta)}\right)\\\label{k}
K^+_{12}(u)&=&K^+_{21}(u)=-\sinh{(2u+2\eta)}
\end{eqnarray}
Herein we have introduced the parameter $c = \log{\left((1-U^2/4)/U\right)}$ for the impurity model, $U$ being the strength of coupling of the impurity to the liquid, or $c=\log{\left(4t/(1-4t^2)\right)}$ for the Weak -Tunnelling model. Let us denote the latter by $c_t$ when a distinction is required. The logarithmic dependence on the bare coupling constant will be important later when considering thermodynamic quantities, we will see that it leads to generation of a scale with power law dependence on the bare parameters in \eqref{H}. In addition we have also introduced an inhomogeneity parameter $\theta$ which will enable us to relate the monodromy matrix  to $Z$or $Z_t$.  Using the definitions we construct the monodromy matrix,
\begin{eqnarray}\nonumber
\Xi_0(u)=\mathcal{C} K^+(u)\mathcal{R}_{01}(u+\theta/2)\dots \mathcal{R}_{0N}(u+\theta/2)\\\label{Xi}
\times K^-(u)\mathcal{R}_{0N}(u-\theta/2)\dots \mathcal{R}_{01}(u-\theta/2) \label{C}
\end{eqnarray}
with $\mathcal{C}=\frac{-\beta e^{-\eta}}{\sinh{\theta}\sinh{\frac{3\theta}{2}}}$ and $\beta\to\beta_t$ for the Weak-Tunnelling model. An auxiliary space indexed by $0$, very useful for a convenient formulation of the problem, has been introduced.  The form of \eqref{Xi} is similar to that of the XXZ model with two boundaries described by $K^+$ and $K^-$.   The transfer matrix is given by the trace over this auxiliary space, 
\begin{eqnarray}
t(u)&=&\text{Tr}_0\,\Xi(u).
\end{eqnarray}
The judicious choice of boundary matrices means that the transfer matrices commute for differing spectral parameter, $[t(u),t(v)]=0$ \cite{Sklyannin} and by expanding in powers of $u$  a set of operators which commute with $t(v)$ is generated. This proves the integrability of the transfer matrix.

 We now return to our original problem, the diagonalization of $Z$. The choice of \eqref{R} and \eqref{K}-\eqref{k} as well as the dependence of the monodromy matrix on $\theta$ means that we can relate this to the transfer matrix. In particular, setting $u=\theta/2$ we have,
\begin{equation}
Z=\lim_{\theta\rightarrow\infty}t(\theta/2).
\end{equation} 
and similarly $Z_t$ with the appropriate replacements. What we have shown, therefore, is that determining the spectrum of $Z$ or $Z_t$ is related to that of the open XXZ chain with prescribed inhomogeneities, boundaries and twists. In addition we have established the integrability of both the Kane-Fisher impurity and Weak-Tunnelling models. 

At this point  the QIS method ceases to be of use.  The reason for this is the non diagonal nature of the boundary matrices means that there is no proper reference state upon which to build the eigenstates of $t(u)$ and determine the eigenvalues. This can be circumvented by means of the newly developed ODBA approach which  utilises certain algebraic properties of the transfer matrix to completely determine its eigenvalues in terms of an inhomogeneous T-Q relation. The eigenvalue is parametrised by Bethe roots, $\mu_j$ which are fixed by the Bethe equations.  The states can then also be recovered by means of separation of variables \cite{SovODBA}. Presently we are only interested in eigenvalues of $t(u)$ and so postpone any discussion of the states to future work. 

The transfer matrix $t(u)$ has previously been considered in \cite{ODBAXXZ} wherein the eigenvalues, $\Lambda(u)$, and the Bethe equations were determined. Inserting \eqref{K}-\eqref{k} and \eqref{C} into their results we find for $N$ even,
\begin{eqnarray}\nonumber
\Lambda(\theta/2)=-4i\beta e^{i\phi}\frac{\sinh{(\theta-2i\phi)}\cosh{(c)}\cosh{(\theta/2)}}{\sinh{(\theta-i\phi)}\sinh{\theta}}\\\label{Lambda}
\times\cosh{(\theta/2-i\Phi)}\prod^N_j\frac{\sinh{(\theta/2-\mu_j+i\phi)}}{\sinh{(\theta/2+\mu_j-i\phi)}}.
\end{eqnarray}
%with $\mu_j$ being the Bethe parameters. 
We have restricted ourselves to $u=\theta/2$ since we are only interested in determining $e^{-ikL}=\lim_{\theta\to\infty} \Lambda(\theta/2)$.  In addition we obtain the Bethe equations,
\begin{widetext}
\begin{eqnarray}\nonumber
\frac{\left[\cosh{\left(i(N+1)\phi+c+i\pi/2+i\Phi-\theta/2+2\sum^N_{j=1}\mu_j\right)}-1\right]\sinh{(2\mu_j-i\phi)}\sinh{(2\mu_j-2i\phi)}}{2i\cosh{(\mu_j+c+\theta/2-i\phi)}\cosh{(\mu_j-i\phi)}\cosh{(\mu_j-i\phi+i\Phi)}\sinh{(\mu_j-\theta-i\phi)}}\\\label{BAE}
=\prod^N_{l=1}\frac{\sinh{(\mu_j+\mu_l-i\phi)}\sinh{(\mu_j+\mu_l-2i\phi)}}{\sinh{(\mu_j+\theta/2-i\phi)}\sinh{(\mu_j-\theta/2-i\phi)}}
\end{eqnarray}
\end{widetext} 
along with the selection rules $\mu_j\neq\mu_k$ and $\mu_j\neq\mu_k+i\phi$. These selection rules are analogous to the exclusion principle in other Bethe Ansatz problems \cite{Korepin}. Upon taking the limit, $\theta\to\infty$ \eqref{Lambda} and \eqref{BAE} completely determine the spectrum of $Z$. Prior to doing so we should consider the dependence of $\mu_j$ on $\theta$. The dependence of  the Bethe parameters   on   the inhomogeneity $\theta$  follows from the form of \eqref{Lambda} and \eqref{BAE}  with half the roots scaling  as $-\theta/2$ while the other half go as $\theta/2$. This is also the case for $N$ odd,  as $N+1$ Bethe parameters are required by the ODBA solution\cite{ODBAXXZ}. We separate out the $\theta$ dependent part and introduce two sets of Bethe parameters $\{  \lambda_j, \nu_j    \}$, 
\begin{equation}\label{mu}
\mu_j=\begin{cases} \lambda_j+i\phi/2+\theta/2 &
\text{if}j\leq \frac{N}{2}\\
 -\nu_{j-N/2}+i\phi/2-\theta/2&
\text{if}j>\frac{N}{2}.
\end{cases}
\end{equation}
The validity of this assumption will be checked by recovering the Luttinger liquid spectrum when the impurity is removed. Inserting \eqref{mu} into \eqref{Lambda} the eigenvalues become
\begin{equation}\label{ee}
e^{-ikL}=\frac{-e^{-i\Phi}}{\alpha}\prod^{N/2}_j\frac{\sinh{(\lambda_j-i\phi/2)}}{\sinh{(\nu_j+i\phi/2)}}e^{-\lambda_j+\nu_j+i\phi}.
\end{equation}
Two sets of Bethe equations for $\lambda_j$ and $\nu_j$ are obtained from \eqref{BAE} and \eqref{mu},
\begin{eqnarray}\nonumber
&&\sinh^N{(\lambda_j-i\phi/2)}=-e^{-2\lambda_j-i\phi+2c+2i\Phi}e^{2\sum_k(2\lambda_k-\nu_k)}\\\label{LBAE}
&&~~~~~~~~\times\prod^{N/2}_k\sinh{(\lambda_j-\nu_k)}\sinh{(\lambda_j-\nu_k-i\phi)}\\\nonumber
&&\sinh^N{(\nu_j+i\phi/2)}=\frac{2i\cosh{(c-\nu_j-i\phi/2)}}{e^{\nu_j-c+i\phi/2}}e^{2\sum_k\lambda_k}\\\label{NBAE}
&&~~~~~~~~\times\prod^{N/2}_k\sinh{(\nu_j-\lambda_k)}\sinh{(\nu_j-\lambda_k+i\phi)}
\end{eqnarray}
 with the selection rules now reading $\lambda_j\neq\nu_k,~\lambda_j\neq\lambda_k,~\nu_j\neq\nu_k$. The complexity of both the eigenvalues and Bethe equations is a common feature of models solved by ODBA and accordingly makes them more difficult to treat. However we can gain some insight as to the structure of the solutions by considering the case of weak or vanishing impurity strength $U \to 0$. This will also serve as  a check on \eqref{mu} by correctly reproducing the spectrum of the Luttinger Liquid.  In this limit the impurity parameter, $c\to\infty$,  blows up. Inserting this in \eqref{LBAE}, \eqref{NBAE} we see that the solutions  are either $\lambda_j=\nu_j$ or $\lambda_j=\nu_j+i\phi$. In terms of the original parameters these are $\mu_{j+N/2}=-\mu_j+i\phi$ or $\mu_{j+N/2}=-\mu_j+2i\phi$. This leaves half the parameters, $\mu_j, ~j\leq N/2$ undetermined. To fix these remaining $\mu_j$, we return to the expression for $\Lambda(u)$ as given by \cite{ODBAXXZ} and assume there are $M$ pairs such that $\mu_{j+N/2}=-\mu_j+i\phi$   while the other $N/2-M$ are of the form $\mu_{j+N/2}=-\mu_j+2i\phi$. Upon taking $c\to\infty$ we find that the $N/2-M$ latter pairs decouple and we are left with a T-Q relation in terms of $M$ parameters $\mu_j$ (see Appendix B). From this we derive the eigenvalues
\begin{eqnarray}\label{E}
e^{-ikL}=e^{Mi\phi-i\Phi}\prod^M_{j=1}\frac{\sinh{(\lambda_j-i\phi/2)}}{\sinh{(\lambda_j+i\phi/2)}}.
\end{eqnarray}
The Bethe equations are similar to those of the XXZ model,
\begin{eqnarray}\nonumber
\frac{\sinh^N{(\lambda_j-i\phi/2)}}{\sinh^N{(\lambda_j+i\phi/2)}}&=&e^{i(N-2M)\phi+2i\Phi}\\\label{llBae}
&&\times \prod^M_{k\neq j}\frac{\sinh{(\lambda_j-\lambda_k-i\phi)}}{\sinh{(\lambda_j-\lambda_k+i\phi)}}. 
\end{eqnarray}
The extra phase factor in the Bethe equations will not change the structure of the solutions which are either real or form strings in the Thermodynamic limit \cite{Takahashi} for $-\pi\leq\phi\leq\pi$. It is however, crucial in obtaining the correct energy of the Luttinger liquid. 
 Combining \eqref{E} and \eqref{llBae} we obtain, 
\begin{eqnarray}\nonumber
E=\frac{2\pi}{L}\sum^N_kn_k-\frac{2\pi}{L}\sum_j^MI_j-\frac{2M(N-M)}{L}\phi\\\label{lle} +\frac{\Phi}{L}(N-2M).
\end{eqnarray}
Here $n_k$ and $I_j$ are the quantum numbers associated to the charge and chiral degrees of freedom. The last term is recognisable as $-\mathcal{A}(N_+-N_-)$.  This validates our choice of \eqref{mu}. 

Before proceeding to a study of the impurity thermodynamics we should note that  strings represent gapless excitations of the Luttinger liquid and  their structure  depends heavily on the strength of the interaction. While we have successfully diagonalised the model for all $\phi$ and $U\geq0$, for clarity we hereafter restrict ourselves to the simplest structure and take $|\phi|=\pi/\nu$ with $\nu >2$ an integer. This then fixes the allowed string lengths and parities. Common to other integrable models we can have $j$-strings
\begin{equation}
\lambda^{(j,l)}=\lambda^j+i(2j+1-l)\phi/2,
\end{equation}
for $j=1\dots,\nu-1$. These are said to have parity $v_j=1$. In addition to these we may also have strings of negative parity, $v_\nu=-1$ which are centred on the $i\pi/2$ axis. As a consequence of our choice of $\phi$, however only $1$-strings of negative parity are allowed, 
\begin{equation}
\lambda^\nu_\alpha+i\pi/2.
\end{equation}
Once again these represent bulk excitations and so will not be affected by the introduction of a local impurity. Our choice of scattering Bethe has dictated these as the appropriate excitations of the bulk which diagonalise the impurity.

The formal similarity between   the  Bethe Ansatz equations of the XXZ systems with boundaries and the impurity Luttinger system arises from the analogy of spin degrees of freedom in the first and the chiral degrees of freedom in the second system, though their dynamics is of course very different. We note that for the XXZ with generic boundary fields the residual $U(1)$ spin symmetry is broken by the off diagonal elements of the boundary matrices and it is this that necessitates the use of the ODBA. For the Luttinger liquid we also have a $U(1)$ symmetry (with charge $N_+-N_-$) which is why we are led to taking the XXZ R-matrix while the inclusion of the impurity breaks this and forces us to adopt the ODBA.

\section{Thermodynamics}
Having shown how the spectra of $Z$ and $Z_t$ are described by \eqref{ee}, \eqref{LBAE} and \eqref{NBAE}  we  determine from it the spectrum of $H$ and $H_{WT}$ and proceed to study their thermodynamic behaviour. In particular we calculate the free energy and entropy of the impurity and tunnel junction. In doing so we are interested in impurity effects but not finite size effects. As a result we will lose sensitivity to the influence of the flux $\Phi$ \cite{Shastri}. In the following we set $\Phi$ to zero and  will address  transport  properties through the Kubo formula. 

Dealing directly with \eqref{LBAE} and \eqref{NBAE} is arduous  due to their non standard form but  methods have been developed to extract physical quantities in the thermodynamic limit \cite{BdryEnergy}. Here we will adopt a different approach.  We have just seen that for $c\to\infty$ the eigenvalues and Bethe equations are given by \eqref{E} and \eqref{llBae}. For large but finite $c$, corresponding to $U\ll 1$ the form of these equations are modified by an impurity term which is necessarily of the order $1/N$. Indeed we know that any bulk properties cannot be modified by introducing an impurity. Thus, we make the assumption that the Bethe parameters are either real, form strings of positive parity such that
\begin{eqnarray}
\text{Im}\{\lambda^{(j,l)}\}=\text{Im}\{\nu^{(j,l)}\}=(2j+1-l)\phi/2
\end{eqnarray}
or negative parity Im$\{\lambda_j\}=$Im$\{\nu_j\}=\pi/2$ in the thermodynamic limit 
or come in pairs $\text{Im}\{\lambda_j-\nu_j\}=\phi$. 
  
Proceeding from this assumption we can derive the continuous form of the  Bethe Ansatz equations (BAE).  The result is that the distributions for the $j$-strings and holes,  $\rho_j(x)$ and holes $\rho_j^h(x)$ \cite{Takahashi} satisfy,
\begin{eqnarray}\label{TBA}
Na_j(x)+b_j(x)=\rho_j(x)+\rho_j^h(x)+\sum_k^\nu A_{jk}*\rho_k(x)\; \; \;\;\\
Na_\nu(x)+b_\nu(x)=-\rho_\nu(x)-\rho_\nu^h(x)+\sum_k^\nu A_{\nu k}*\rho_k(x)\; \; \;
\end{eqnarray} 
where we define:
\begin{eqnarray}
a_j(x)&=&\frac{1}{2\pi}\d{}{x}p(x,n_j,v_j)\\
A_{jk}(x)&=&\frac{1}{2\pi}\d{}{x}\Theta_{jk}(x)\\
b_j(x)&=&-\frac{1}{4\pi}\d{}{x}p(x-c/\phi,n_j,-v_j)
\end{eqnarray}
with
\begin{eqnarray}
p(x,n_j,v_j)=2v_j\arctan{\left((\cot{n_j\phi/2})^{v_j}\tanh{\phi x}\right)} \;\;\;\;\;
 \\
\Theta_{jk}(x)=p(x,|n_j-n_k|,v_jv_k)+p(x,n_j+n_k,v_jv_k) \nonumber\\
 +2\sum_q p(x,|n_j-n_k|+2q,v_jv_k)
 \end{eqnarray}
and $*$ denoting a convolution $f*g(x)=\int \mathrm{d} y\,f(x-y)g(y)$.
The change in sign for the $v=-1$ roots arises because $p_j(x,n_j,v_j)$ changes from monotonically increasing to decreasing when $v_j=1\to v_j=-1$. In order to have $\rho_\nu(x)\geq 0$ we need to introduce the sign. The energy in terms of these string configurations is
\begin{equation}\label{Estring}
E=-\sum_{j=1}^\nu D\int\rho_j(x)\left(p(x,n_j,v_j)+\theta(v_j)\pi\right).
\end{equation}

The form of the Bethe equations is very similar to the that of the anisotropic Kondo model (AKM). Indeed if we change the parity of the impurity terms, $b_j(x)$, from $-1$ to $1$ so that it is now $a_j(x)$ we recover the equations for the AKM  with zero external field \cite{TWAKM}. The change in the parity of the impurity term can be understood by noticing the impurity we presently consider is not merely a particle at a fixed location but introduces a new aspect, the mixing of the left and right movers this is in contrast to the Kondo model or AKM. In addition the change in parity ensures that if the non interacting limit is taken, $\phi\to 0$, the impurity term vanishes and the distributions are those of free fermions.

We now proceed to construct the free energy by means of the Yang-Yang approach and its generalisation.  The approach is well known and we just provide the main steps. The free energy, $F=E-TS$,  where $E$ is given by \eqref{Estring} and  $S=\sum_j\int \left[(\rho_j+\rho_j^h)\log{(\rho_j+\rho_j^h)}-\rho_j\log{(\rho_j)}-\rho^h_j\log{(\rho^h_j)}\right]$ is the entropy associated to the distributions, is minimised with respect to $\rho_j$ which are solutions of the BAE. 
The result of this minimisation gives the thermodynamic Bethe ansatz equations (TBA), 
\begin{eqnarray}\nonumber
\log{\eta_j(x)}=s*\log{(1+\eta_{j+1}(x))(1+\eta_{j-1}(x))}~~~~~~~~\\\label{gt}
+\delta_{j,\nu-2}s*\log{(1+\eta^{-1}_\nu(x))}-\delta_{j,1}\frac{2D}{T}\arctan{e^{\pi x}}\\
\log{\eta_{\nu-1}(x)}=s*\log{(1+\eta_{\nu-2}(x))}=-\log{\eta_{\nu}(x)}~~~
\end{eqnarray}
with $\eta_j(x)=\rho_j^h(x)/\rho_j(x)$, $s(x)=\frac{1}{2\cosh{\pi x}}$. The density $D=\frac{N}{L}$ plays also the role bandwidth up to a factor of $\pi$ for the linear spectrum: setting $k_F=0$  the ground state is filled down to  $-N \frac{2\pi}{L}$.

Having taken the thermodynamic limit and derived the TBA equations we proceed to take the scaling limit  to obtain universal quantities, eliminating any dependence on $D$.  As we shall see the the model generates an energy scale $T_{KF}$ which will be held fixed as $D \to \infty$. Thus high and low temperature regimes will be defined with respect to $T_{KF}$  and always small compared to $D$. With this in mind we introduce the universal functions \cite{TWAKM}, 
\begin{eqnarray}\label{U}
\varphi_j(x)&=&\frac{1}{T}\log{\big(\eta_j(x+\frac{1}{\pi}\log{\frac{T}{D}})\big)}.
\end{eqnarray}
Inserting these into \eqref{gt}  and approximating the driving term, $-\frac{2D}{T}\arctan{\exp{\pi (x+\frac{1}{\pi}\log{\frac{T}{D}}}}\simeq -2e^{\pi x}$, an approximation valid since only this range of values contributes to $\eta_1(x)$, we obtain the universal (or scaling) form of the TBA equations,
\begin{eqnarray}\nonumber
\varphi_j(x)&=&s*\log{(1+e^{\varphi_{j-1}(x)})(1+e^{\varphi_{j+1}(x)})}\\\label{UTBA}
&&-\delta_{j,1}2e^{\pi x},~~~j<\nu-2\\\nonumber
\varphi_{\nu-2}(x)&=&s*\log(1+e^{\varphi_{\nu-1}(x)})(1+e^{\varphi_{\nu-3}(x)})\\
&&~~~~~~~~~~~~~~~~~~~~~~~~\times(1+e^{-\varphi_\nu (x)}),\\
\varphi_{\nu-1}(x)&=&s*\log{(1+e^{\varphi_{\nu-2}(x)})}=-\varphi_{\nu}(x).
\end{eqnarray}
The free energy can then be written as:
\begin{eqnarray}
F=F^{LL}+F^{i}
\end{eqnarray}
with  $F^{LL}=E_0-T N\int s(x)\log{(1+\exp{\varphi_1(x)})}$  being the bulk contribution ($E_0$ the ground state energy)  which  the impurity contribution is,
\begin{eqnarray}\label{F}
F^{i}=-T\int\mathrm{d} x\, s(x+\frac{1}{\pi}\log\frac{T}{T_{KF}})
\log{(1+e^{\varphi_{\nu-1}(x)})}.
\end{eqnarray}
We note here the appearance of a scale $T_{KF}=De^{\pi c/\phi}$ which has been generated by the model. We will measure all temperatures relative to this scale and can obtain universal results by keeping $T_{KF}$ fixed while taking $D\to\infty$. In terms of the original parameters of the Hamiltonian this is
\begin{eqnarray}
T_{KF}
&=&D\left(\frac{U}{1-U^2/4}\right)^{\frac{\pi}{2\arctan g}}.
\end{eqnarray}
This scale is power law in the interaction strength which matches predictions made by Renormalisation Group techniques \cite{KF}. Having identified the scale we can determine the dependence of the impurity strength on the cutoff $D$.  The behaviour depends on the sign of the interaction strength. For repulsive interactions $g>0$,
\begin{equation}
U(D)\sim \left(\frac{T_{KF}}{D}\right)^{\frac{2\arctan g}{\pi}}
\end{equation}
which show $U\to 0$ as $D\to \infty$,  or running the argument backwards, indicating the strengthening of the impurity at small energy scales as $D$ is decreased. In contrast, for attractive interactions the $U(D)$ grows with the scale signifying the healing of the system at low energy. 

Likewise, the Weak-Tunnelling Hamiltonian also generates a scale $T_{WT}=De^{\pi c_t/\phi}$. The complementary nature of these models is exposed when written in the bare parameters,
\begin{eqnarray}
T_{WT}=D\left(\frac{4t}{1-4t^2}\right)^{-\frac{\pi}{2\arctan g}}.
\end{eqnarray} 
The change in the sign of the exponent causes the tunnelling parameter to run oppositely to the impurity strength. The two systems thus become disjoint when the interactions are repulsive and completely joined for attractive interactions at low energies. 

Any thermodynamic calculations are valid only when the generated scale is less than the cutoff. Accordingly we are hereafter restricted to the repulsive regime of the impurity model and the attractive regime for the Weak-Tunnelling Hamiltonian. We will only present the former but the latter is similar with the appropriate replacement of the scale.

 Having taken the scaling limit  we turn now to study the universal  temperature dependence of the free energy. It requires the full solution of the TBA equations which can be achieved only numerically. Here we shall consider the high $T \gg T_{KF} $ and low temperature  $T \ll T_{KF}$  limits  and leave the study of the crossover to a later publication.
 The free energy is given in terms of $\varphi_{\nu-1}$ which is coupled to all other $\varphi_j$ but  still  we can obtain some results for high and low temperature. At $T\gg T_{KF}$ the integral in \eqref{F} is dominated by the behaviour at $x\to-\infty$, in this limit the driving term drops out of \eqref{gt} and the solutions are constants. Denoting $e^{\varphi_j(-\infty)}=x_j$, we get,
\begin{eqnarray}\label{high}
x_j=(j+1)^2-1,~~~~x_{\nu-1}=\nu-1=1/x_\nu.
\end{eqnarray}
Similarly for low $T\ll T_{KF}$ we look for solutions at $x\to\infty$. This time we denote $e^{\varphi_j(\infty)}=y_j$ and find
\begin{equation}\label{low}
y_j=j^2-1,~~~y_{\nu-1}=\nu-2=1/y_{\nu}.
\end{equation}

Using the expression for the free energy along with~\eqref{low} and \eqref{high}   we can calculate impurity free energy near the UV and IR fixed points,
\begin{eqnarray}
F^i_{UV}=\frac{T}{2}\log{(\nu)},~~F^i_{IR}=\frac{T}{2}\log{(\nu-1)}
\end{eqnarray} The difference in the impurity entropy between  fixed points,
\begin{equation}\label{Ent}
S^i_{UV}-S^i_{IR}=\frac{1}{2}\log{\frac{\nu}{\nu-1}}
\end{equation}
 shows the usual decrease  as the system flows from the UV to the IR fixed points  (a flow from weak to strong coupling regime for repulsive interactions), a decrease which in the language of the renormalisation group corresponds the  degrees of freedom that were  integrated out.
The form of this result agrees with the values calculated for the boundary terms in both the boundary Sine-Gordon model \cite{FSW} as well as XXZ with parallel boundary fields \cite{deSa}, however the degrees of freedom  as well as the interpretation of $\nu$ are different in those cases. 

 We now consider the corrections  to the asymptotic limits  
 \eqref{high}  and  \eqref{low}  which can also be calculated  \cite{deSa}.  The corrections yield  the specific heat  which is found to scale as,
\begin{eqnarray}\label{c}
C(T\ll T_{KF})&\sim & \left(\frac{T}{T_{KF}}\right)^\frac{2}{\nu-1}\\
C(T\gg T_{KF})&\sim &\left(\frac{T_{KF}}{T}\right)^\frac{2}{\nu}
\end{eqnarray}
indicating that both the  strong and weak coupling fixed point are Non-Fermi Liquid in nature. 

Using arguments from boundary conformal field theory \cite{AL} we can identify the leading irrelevant operators at both fixed points and thus determine the scaling of the conductance as given by Kubo's formula.  At low temperature  the conductance vanishes  as $G \sim T ^\frac{2}{\nu-1}$ corresponding to the effective increase of the strength of the impurity $U$ as $D$ is decreased noted earlier. Thus the low temperature physics is governed by strong coupling  Hamiltonian where the wire is cut by the impurity and for which the Weak-Tunnelling model is the starting point. At  high temperatures, in addition to the wire conductance  $G_0= K e^2/h$, with $K=(\nu -1)/\nu$ for our choice of $\phi$,  we have  the impurity correction $ G \sim T^{-\frac{2}{\nu}}$, its vanishing at high temperatures  corresponding to the healing of the wire  \cite{KF}. We expect similar results to be obtained from finite size calculations on a ring threaded by flux $\Phi$.

\section{elementary excitations}
In the previous section we derived the impurity thermodynamics of both  the Kane-Fisher impurity model and Weak-Tunnelling model with spin isotropic bulk interaction. Here we will discuss the elementary excitation of the models, which we call {\it chirons} owing to their origin in the chiral degrees of freedom. 
The ground state of the system contains only real roots whose distribution is governed by the $j=1$ equation of \eqref{TBA} with the $\rho^h_1(x)=\rho_j(x)=0$ for $j>1$. Excitations above this ground state are obtained by adding holes in this distribution. The chiron energy, $\varepsilon=2D\arctan{e^{\pi x^h}}$, appears as the diving term in the TBA equations \eqref{gt} with $x^h$ being the position of the hole in the distribution.   

Using the method of \cite{4lectures} we can determine their phase shift as they scatter past the impurity. To do this we note that in the absence of the impurity the chiron energy should take on values $2\pi I^h/L$ (See Eq.\eqref{lle}). The $1/L$ deviation of $\varepsilon$ from  this value gives the chiron- impurity phase shift. Up to an overall constant phase the impurity S-matrix is  
\begin{eqnarray}
S^{c,i}(\varepsilon)&=&e^{i\Delta^{c,i}(\frac{1}{\pi}\log{(\varepsilon/T_{KF})})},\\
\Delta^{c,i}(x)&=&\int\frac{\mathrm{d}\omega}{8\pi i\omega}\frac{\tanh{(\omega/2)}}{\sinh{\left((\pi/\phi-1)\omega/2\right)}}e^{i\omega x} \nonumber
\end{eqnarray}
This is valid for $\pi/\phi$ being an arbitrary rational number between $0$ and $1$. We see that the phase shift is non trivial at both low and high energies as both IR and UV fixed points are non trivial. This is to be compared with bare electrons which are perfectly transmitted at high energies and reflected at low energy. 

Adding two holes to the ground state distribution allows us to calculate the chiron-chiron  phase shift in the same manner,
\begin{eqnarray} 
S^{c,c}(\varepsilon_1,\varepsilon_2)&=&e^{i\Delta^{c,c}(\varepsilon_1-\varepsilon_2)}, \\  
\Delta^{c,c}(x)&=&\int\frac{\mathrm{d}\omega}{4\pi i\omega}\frac{\sinh{((\pi/\phi-2)\omega/2)}e^{i\omega x} }{\cosh{(\omega/2)}\sinh{\left((\pi/\phi-1)\omega/2\right)}}\nonumber
\end{eqnarray}  
With $\varepsilon_j$ the energies of the two chirons. The full physical spectrum is thus built up by adding holes and strings to the ground state distribution. The interpretation of the strings is commented on below.

 We now turn to discuss the relation between our approach with the bootstrap approach where the spectrum of the Hamiltonian and the various S-matrices are postulated on the basis of integrability properties.  It is known that the  impurity model without spin is related via bosonisation and folding procedures to the massless limit of the boundary Sine-Gordon model.  Its spectrum is taken to consist of Solitons, anti-Solitons and their bound states known as Breathers. The dressed S-matrices, derived via the bootstrap method of \cite{Zam}, are non diagonal for generic interaction strength and calculating thermodynamic quantities leads to an equation similar in structure to \eqref{PBC}. For special values of the interaction however, the bulk scattering becomes diagonal and the computations simplify considerably, the right hand side becoming a mere phase. The inclusion of spin in this method is more complicated and is only achieved in certain interaction regimes \cite{Qwire}.

In contrast the present method constitutes a bottom up approach. We have diagonalised the actual  quantum Hamiltonian with spin for all values of the interaction, our restriction to $\phi=\pi/\nu$ is purely for the convenience of its simplified string structure. It is in this parameter regime where the TBA and free energy in both approaches coincide allowing us to identify the first $\nu-2$ string distributions with Breathers and the last two with symmetric and anti-symmetric combinations of a  Soliton and anti-Soliton.

\section{Conclusions}
In this paper we have solved exactly two related Hamiltonians, a spin isotropic Luttinger liquid coupled to an impurity or a tunnel junction with arbitrary boundary conditions. This was achieved via a new type of coordinate Bethe ansatz that incorporates the reflecting and transmitting properties of the impurity in conjunction with the Off Diagonal Bethe Ansatz. We found that determining the spectrum is equivalent to an analogous problem for an open XXZ chain with one boundary corresponding to the impurity and the other the boundary condition. The thermodynamics was then studied and it was shown that a scale is naturally generated by both models such that the impurity strength and tunnelling parameter run oppositely confirming the duality of the models. The impurity free energy for the simplest interaction regime was calculated and was seen to coincide with that obtained in \cite{FSW} for the case without spin. The diagonalisation of the 
model allows us to view the system as a gas of excitations in the chiral degrees of freedom, chirons, which scatter with a pure phase off the impurity. 

The methods presented herein, we believe to be quite general and provide a template for solving other impurity models with interacting bulk. Indeed the coordinate Bethe ansatz is readily applied to the model with spin anisotropic interaction and  with a Kondo impurity.  Moreover the formulation naturally allows for arbitrary boundary conditions to be imposed allowing for the potential to study the effects of impurities on mesoscopic rings with arbitrary flux \cite{Meso}.

\acknowledgements{This research was supported by NSF grant DMR 1410583. We are grateful to Sung-Po Chao, Yashar Komijani and Giuseppe Mussardo  for useful discussions.}

\bibliography{mybib}

\begin{thebibliography}{27}
\expandafter\ifx\csname natexlab\endcsname\relax\def\natexlab#1{#1}\fi
\expandafter\ifx\csname bibnamefont\endcsname\relax
  \def\bibnamefont#1{#1}\fi
\expandafter\ifx\csname bibfnamefont\endcsname\relax
  \def\bibfnamefont#1{#1}\fi
\expandafter\ifx\csname citenamefont\endcsname\relax
  \def\citenamefont#1{#1}\fi
\expandafter\ifx\csname url\endcsname\relax
  \def\url#1{\texttt{#1}}\fi
\expandafter\ifx\csname urlprefix\endcsname\relax\def\urlprefix{URL }\fi
\providecommand{\bibinfo}[2]{#2}
\providecommand{\eprint}[2][]{\url{#2}}

\bibitem[{\citenamefont{Giamarchi}(2003)}]{TG}
\bibinfo{author}{\bibfnamefont{T.}~\bibnamefont{Giamarchi}},
  \emph{\bibinfo{title}{Quantum Physics in One Dimension}}, International
  Series of Monographs on Physics (\bibinfo{publisher}{Clarendon Press},
  \bibinfo{year}{2003}), ISBN \bibinfo{isbn}{9780198525004}.

\bibitem[{\citenamefont{Kane and Fisher}(1992)}]{KF}
\bibinfo{author}{\bibfnamefont{C.~L.} \bibnamefont{Kane}} \bibnamefont{and}
  \bibinfo{author}{\bibfnamefont{M.~P.~A.} \bibnamefont{Fisher}},
  \bibinfo{journal}{Phys. Rev. B} \textbf{\bibinfo{volume}{46}},
  \bibinfo{pages}{15233} (\bibinfo{year}{1992}),
  \urlprefix\url{http://link.aps.org/doi/10.1103/PhysRevB.46.15233}.

\bibitem[{\citenamefont{Chang}(2003)}]{LLRMP}
\bibinfo{author}{\bibfnamefont{A.~M.} \bibnamefont{Chang}},
  \bibinfo{journal}{Rev. Mod. Phys.} \textbf{\bibinfo{volume}{75}},
  \bibinfo{pages}{1449} (\bibinfo{year}{2003}),
  \urlprefix\url{http://link.aps.org/doi/10.1103/RevModPhys.75.1449}.

\bibitem[{\citenamefont{Safi and Saleur}(2004)}]{circuit}
\bibinfo{author}{\bibfnamefont{I.}~\bibnamefont{Safi}} \bibnamefont{and}
  \bibinfo{author}{\bibfnamefont{H.}~\bibnamefont{Saleur}},
  \bibinfo{journal}{Phys. Rev. Lett.} \textbf{\bibinfo{volume}{93}},
  \bibinfo{pages}{126602} (\bibinfo{year}{2004}),
  \urlprefix\url{http://link.aps.org/doi/10.1103/PhysRevLett.93.126602}.

\bibitem[{\citenamefont{Bloch et~al.}(2008)\citenamefont{Bloch, Dalibard, and
  Zwerger}}]{IBRMP}
\bibinfo{author}{\bibfnamefont{I.}~\bibnamefont{Bloch}},
  \bibinfo{author}{\bibfnamefont{J.}~\bibnamefont{Dalibard}}, \bibnamefont{and}
  \bibinfo{author}{\bibfnamefont{W.}~\bibnamefont{Zwerger}},
  \bibinfo{journal}{Rev. Mod. Phys.} \textbf{\bibinfo{volume}{80}},
  \bibinfo{pages}{885} (\bibinfo{year}{2008}),
  \urlprefix\url{http://link.aps.org/doi/10.1103/RevModPhys.80.885}.

\bibitem[{\citenamefont{{Krinner} et~al.}(2015)\citenamefont{{Krinner},
  {Stadler}, {Husmann}, {Brantut}, and {Esslinger}}}]{Coldatom}
\bibinfo{author}{\bibfnamefont{S.}~\bibnamefont{{Krinner}}},
  \bibinfo{author}{\bibfnamefont{D.}~\bibnamefont{{Stadler}}},
  \bibinfo{author}{\bibfnamefont{D.}~\bibnamefont{{Husmann}}},
  \bibinfo{author}{\bibfnamefont{J.-P.} \bibnamefont{{Brantut}}},
  \bibnamefont{and}
  \bibinfo{author}{\bibfnamefont{T.}~\bibnamefont{{Esslinger}}},
  \bibinfo{journal}{\nat} \textbf{\bibinfo{volume}{517}}, \bibinfo{pages}{64}
  (\bibinfo{year}{2015}), \eprint{1404.6400}.

\bibitem[{\citenamefont{Husmann et~al.}(2015)\citenamefont{Husmann, Uchino,
  Krinner, Lebrat, Giamarchi, Esslinger, and Brantut}}]{esslinger}
\bibinfo{author}{\bibfnamefont{D.}~\bibnamefont{Husmann}},
  \bibinfo{author}{\bibfnamefont{S.}~\bibnamefont{Uchino}},
  \bibinfo{author}{\bibfnamefont{S.}~\bibnamefont{Krinner}},
  \bibinfo{author}{\bibfnamefont{M.}~\bibnamefont{Lebrat}},
  \bibinfo{author}{\bibfnamefont{T.}~\bibnamefont{Giamarchi}},
  \bibinfo{author}{\bibfnamefont{T.}~\bibnamefont{Esslinger}},
  \bibnamefont{and} \bibinfo{author}{\bibfnamefont{J.-P.}
  \bibnamefont{Brantut}}, \bibinfo{journal}{Science}
  \textbf{\bibinfo{volume}{350}}, \bibinfo{pages}{1498} (\bibinfo{year}{2015}),
  ISSN \bibinfo{issn}{0036-8075}.

\bibitem[{\citenamefont{{Kinoshita} et~al.}(2006)\citenamefont{{Kinoshita},
  {Wenger}, and {Weiss}}}]{NC}
\bibinfo{author}{\bibfnamefont{T.}~\bibnamefont{{Kinoshita}}},
  \bibinfo{author}{\bibfnamefont{T.}~\bibnamefont{{Wenger}}}, \bibnamefont{and}
  \bibinfo{author}{\bibfnamefont{D.~S.} \bibnamefont{{Weiss}}},
  \bibinfo{journal}{\nat} \textbf{\bibinfo{volume}{440}}, \bibinfo{pages}{900}
  (\bibinfo{year}{2006}).

\bibitem[{\citenamefont{Wang et~al.}(2015)\citenamefont{Wang, Yang, Cao, and
  Shi}}]{ODBA}
\bibinfo{author}{\bibfnamefont{Y.}~\bibnamefont{Wang}},
  \bibinfo{author}{\bibfnamefont{W.-L.} \bibnamefont{Yang}},
  \bibinfo{author}{\bibfnamefont{J.}~\bibnamefont{Cao}}, \bibnamefont{and}
  \bibinfo{author}{\bibfnamefont{K.}~\bibnamefont{Shi}},
  \emph{\bibinfo{title}{{Off-diagonal Bethe ansatz for exactly solvable
  models}}} (\bibinfo{publisher}{Springer}, \bibinfo{address}{Berlin},
  \bibinfo{year}{2015}).

\bibitem[{\citenamefont{{Delfino} et~al.}(1994)\citenamefont{{Delfino},
  {Mussardo}, and {Simonetti}}}]{mussardo}
\bibinfo{author}{\bibfnamefont{G.}~\bibnamefont{{Delfino}}},
  \bibinfo{author}{\bibfnamefont{G.}~\bibnamefont{{Mussardo}}},
  \bibnamefont{and}
  \bibinfo{author}{\bibfnamefont{P.}~\bibnamefont{{Simonetti}}},
  \bibinfo{journal}{Nuclear Physics B} \textbf{\bibinfo{volume}{432}},
  \bibinfo{pages}{518} (\bibinfo{year}{1994}), \eprint{hep-th/9409076}.

\bibitem[{\citenamefont{Brezin and Zinn-Justin}(1966)}]{ZinnJustin}
\bibinfo{author}{\bibfnamefont{E.}~\bibnamefont{Brezin}} \bibnamefont{and}
  \bibinfo{author}{\bibfnamefont{J.}~\bibnamefont{Zinn-Justin}},
  \bibinfo{journal}{Compt. Rend., Ser. B, 263: 671-3(Sept. 12, 1966).}
  (\bibinfo{year}{1966}).

\bibitem[{\citenamefont{{Cao} et~al.}(2013)\citenamefont{{Cao}, {Yang}, {Shi},
  and {Wang}}}]{ODBAXXZ}
\bibinfo{author}{\bibfnamefont{J.}~\bibnamefont{{Cao}}},
  \bibinfo{author}{\bibfnamefont{W.-L.} \bibnamefont{{Yang}}},
  \bibinfo{author}{\bibfnamefont{K.}~\bibnamefont{{Shi}}}, \bibnamefont{and}
  \bibinfo{author}{\bibfnamefont{Y.}~\bibnamefont{{Wang}}},
  \bibinfo{journal}{Nuclear Physics B} \textbf{\bibinfo{volume}{877}},
  \bibinfo{pages}{152} (\bibinfo{year}{2013}), \eprint{1307.2023}.

\bibitem[{\citenamefont{De~Vega and Gonzalez~Ruiz}(1993)}]{DeVega}
\bibinfo{author}{\bibfnamefont{H.}~\bibnamefont{De~Vega}} \bibnamefont{and}
  \bibinfo{author}{\bibfnamefont{A.}~\bibnamefont{Gonzalez~Ruiz}},
  \bibinfo{journal}{Journal of Physics A: Mathematical and General}
  \textbf{\bibinfo{volume}{26}}, \bibinfo{pages}{L519} (\bibinfo{year}{1993}).

\bibitem[{\citenamefont{{Sklyanin}}(1988)}]{Sklyannin}
\bibinfo{author}{\bibfnamefont{E.~K.} \bibnamefont{{Sklyanin}}},
  \bibinfo{journal}{Journal of Physics A Mathematical General}
  \textbf{\bibinfo{volume}{21}}, \bibinfo{pages}{2375} (\bibinfo{year}{1988}).

\bibitem[{\citenamefont{{Zhang} et~al.}(2015)\citenamefont{{Zhang}, {Li},
  {Cao}, {Yang}, {Shi}, and {Wang}}}]{SovODBA}
\bibinfo{author}{\bibfnamefont{X.}~\bibnamefont{{Zhang}}},
  \bibinfo{author}{\bibfnamefont{Y.-Y.} \bibnamefont{{Li}}},
  \bibinfo{author}{\bibfnamefont{J.}~\bibnamefont{{Cao}}},
  \bibinfo{author}{\bibfnamefont{W.-L.} \bibnamefont{{Yang}}},
  \bibinfo{author}{\bibfnamefont{K.}~\bibnamefont{{Shi}}}, \bibnamefont{and}
  \bibinfo{author}{\bibfnamefont{Y.}~\bibnamefont{{Wang}}},
  \bibinfo{journal}{Nuclear Physics B} \textbf{\bibinfo{volume}{893}},
  \bibinfo{pages}{70} (\bibinfo{year}{2015}), \eprint{1412.6905}.

\bibitem[{\citenamefont{{Korepin} et~al.}(1993)\citenamefont{{Korepin},
  {Bogoliubov}, and {Izergin}}}]{Korepin}
\bibinfo{author}{\bibfnamefont{V.~E.} \bibnamefont{{Korepin}}},
  \bibinfo{author}{\bibfnamefont{N.~M.} \bibnamefont{{Bogoliubov}}},
  \bibnamefont{and} \bibinfo{author}{\bibfnamefont{A.~G.}
  \bibnamefont{{Izergin}}}, \emph{\bibinfo{title}{{Quantum Inverse Scattering
  Method and Correlation Functions}}} (\bibinfo{year}{1993}).

\bibitem[{\citenamefont{{Takahashi}}(1999)}]{Takahashi}
\bibinfo{author}{\bibfnamefont{M.}~\bibnamefont{{Takahashi}}},
  \emph{\bibinfo{title}{{Thermodynamics of One-Dimensional Solvable Models}}}
  (\bibinfo{year}{1999}).

\bibitem[{\citenamefont{Shastry and Sutherland}(1990)}]{Shastri}
\bibinfo{author}{\bibfnamefont{B.~S.} \bibnamefont{Shastry}} \bibnamefont{and}
  \bibinfo{author}{\bibfnamefont{B.}~\bibnamefont{Sutherland}},
  \bibinfo{journal}{Phys. Rev. Lett.} \textbf{\bibinfo{volume}{65}},
  \bibinfo{pages}{243} (\bibinfo{year}{1990}),
  \urlprefix\url{http://link.aps.org/doi/10.1103/PhysRevLett.65.243}.

\bibitem[{\citenamefont{{Li} et~al.}(2014)\citenamefont{{Li}, {Cao}, {Yang},
  {Shi}, and {Wang}}}]{BdryEnergy}
\bibinfo{author}{\bibfnamefont{Y.-Y.} \bibnamefont{{Li}}},
  \bibinfo{author}{\bibfnamefont{J.}~\bibnamefont{{Cao}}},
  \bibinfo{author}{\bibfnamefont{W.-L.} \bibnamefont{{Yang}}},
  \bibinfo{author}{\bibfnamefont{K.}~\bibnamefont{{Shi}}}, \bibnamefont{and}
  \bibinfo{author}{\bibfnamefont{Y.}~\bibnamefont{{Wang}}},
  \bibinfo{journal}{Nuclear Physics B} \textbf{\bibinfo{volume}{884}},
  \bibinfo{pages}{17} (\bibinfo{year}{2014}), \eprint{1401.3045}.

\bibitem[{\citenamefont{{Tsvelick} and {Wiegmann}}(1983)}]{TWAKM}
\bibinfo{author}{\bibfnamefont{A.~M.} \bibnamefont{{Tsvelick}}}
  \bibnamefont{and} \bibinfo{author}{\bibfnamefont{P.~B.}
  \bibnamefont{{Wiegmann}}}, \bibinfo{journal}{Advances in Physics}
  \textbf{\bibinfo{volume}{32}}, \bibinfo{pages}{453} (\bibinfo{year}{1983}).

\bibitem[{\citenamefont{{Fendley} et~al.}(1994)\citenamefont{{Fendley},
  {Saleur}, and {Warner}}}]{FSW}
\bibinfo{author}{\bibfnamefont{P.}~\bibnamefont{{Fendley}}},
  \bibinfo{author}{\bibfnamefont{H.}~\bibnamefont{{Saleur}}}, \bibnamefont{and}
  \bibinfo{author}{\bibfnamefont{N.~P.} \bibnamefont{{Warner}}},
  \bibinfo{journal}{Nuclear Physics B} \textbf{\bibinfo{volume}{430}},
  \bibinfo{pages}{577} (\bibinfo{year}{1994}), \eprint{hep-th/9406125}.

\bibitem[{\citenamefont{{de Sa} and {Tsvelik}}(1995)}]{deSa}
\bibinfo{author}{\bibfnamefont{P.~A.} \bibnamefont{{de Sa}}} \bibnamefont{and}
  \bibinfo{author}{\bibfnamefont{A.~M.} \bibnamefont{{Tsvelik}}},
  \bibinfo{journal}{\prb} \textbf{\bibinfo{volume}{52}}, \bibinfo{pages}{3067}
  (\bibinfo{year}{1995}), \eprint{cond-mat/9503031}.

\bibitem[{\citenamefont{Affleck and Ludwig}(1993)}]{AL}
\bibinfo{author}{\bibfnamefont{I.}~\bibnamefont{Affleck}} \bibnamefont{and}
  \bibinfo{author}{\bibfnamefont{A.~W.~W.} \bibnamefont{Ludwig}},
  \bibinfo{journal}{Phys. Rev. B} \textbf{\bibinfo{volume}{48}},
  \bibinfo{pages}{7297} (\bibinfo{year}{1993}),
  \urlprefix\url{http://link.aps.org/doi/10.1103/PhysRevB.48.7297}.

\bibitem[{\citenamefont{Andrei}(1992, cond-mat/9408101)}]{4lectures}
\bibinfo{author}{\bibfnamefont{N.}~\bibnamefont{Andrei}}, in
  \emph{\bibinfo{booktitle}{Series on Modern Condensed Matter Physics - Vol. 6,
  Lecture Notes of ICTP Summer Course}}, edited by
  \bibinfo{editor}{\bibfnamefont{G.~M.} \bibnamefont{S.~Lundquist}}
  \bibnamefont{and} \bibinfo{editor}{\bibfnamefont{Y.}~\bibnamefont{Lu}}
  (\bibinfo{publisher}{World Scientific}, \bibinfo{year}{1992,
  cond-mat/9408101}).

\bibitem[{\citenamefont{{Ghoshal} and {Zamolodchikov}}(1994)}]{Zam}
\bibinfo{author}{\bibfnamefont{S.}~\bibnamefont{{Ghoshal}}} \bibnamefont{and}
  \bibinfo{author}{\bibfnamefont{A.}~\bibnamefont{{Zamolodchikov}}},
  \bibinfo{journal}{International Journal of Modern Physics A}
  \textbf{\bibinfo{volume}{9}}, \bibinfo{pages}{3841} (\bibinfo{year}{1994}),
  \eprint{hep-th/9306002}.

\bibitem[{\citenamefont{Lesage et~al.}(1997)\citenamefont{Lesage, Saleur, and
  Simonetti}}]{Qwire}
\bibinfo{author}{\bibfnamefont{F.}~\bibnamefont{Lesage}},
  \bibinfo{author}{\bibfnamefont{H.}~\bibnamefont{Saleur}}, \bibnamefont{and}
  \bibinfo{author}{\bibfnamefont{P.}~\bibnamefont{Simonetti}},
  \bibinfo{journal}{Phys. Rev. B} \textbf{\bibinfo{volume}{56}},
  \bibinfo{pages}{7598} (\bibinfo{year}{1997}),
  \urlprefix\url{http://link.aps.org/doi/10.1103/PhysRevB.56.7598}.

\bibitem[{\citenamefont{{Eckle} et~al.}(2001)\citenamefont{{Eckle},
  {Johannesson}, and {Stafford}}}]{Meso}
\bibinfo{author}{\bibfnamefont{H.-P.} \bibnamefont{{Eckle}}},
  \bibinfo{author}{\bibfnamefont{H.}~\bibnamefont{{Johannesson}}},
  \bibnamefont{and} \bibinfo{author}{\bibfnamefont{C.~A.}
  \bibnamefont{{Stafford}}}, \bibinfo{journal}{Physical Review Letters}
  \textbf{\bibinfo{volume}{87}}, \bibinfo{eid}{016602} (\bibinfo{year}{2001}),
  \eprint{cond-mat/0010101}.

\end{thebibliography}

\pagebreak
\begin{widetext}
\section*{Appendix A}
In this appendix we derive the eigenvalues \eqref{Lambda} and Bethe equations \eqref{BAE}. First we will review the results of \cite{ODBAXXZ}. They start with the following definitions of R and K matrices,
\begin{eqnarray}
R_{ij}(u)&=&\begin{pmatrix}
\frac{\sinh{u+\eta}}{\sinh{\eta}}&&0&&0&&0\\
0&&\frac{\sinh{u}}{\sinh{\eta}}&&1&&0\\
0&&1&&\frac{\sinh{u}}{\sinh{\eta}}&&0\\
0&&0&&0&&\frac{\sinh{u+\eta}}{\sinh{\eta}}
\end{pmatrix},~~~~K^-(u)=\begin{pmatrix}
K^-_{11}(u)&& K^-_{12}(u)\\
K^-_{21}(u)&&K^-_{22}(u)
\end{pmatrix}\\
K_{11}^-(u)&=&2\left(\sinh{\alpha_-}\cosh{\beta_-}\cosh{u}+\cosh{\alpha_-}\sinh{\beta_-}\sinh{u}\right),\\
K_{22}^-(u)&=&2\left(\sinh{\alpha_-}\cosh{\beta_-}\cosh{u}-\cosh{\alpha_-}\sinh{\beta_-}\sinh{u}\right),\\
K^-_{12}(u)&=&e^{\theta_-}\sinh{2u},~~~~~~~K^-_{21}(u)=e^{-\theta_-}\sinh{2u}
\end{eqnarray}  
Along with these we can define a $K^+(u)=K^-(-u-\eta)$ wherein all subscripts $-$ are replaced by $+$. These then satisfy the reflection equation (RE), dual reflection equation (the RE for $K^+$) and Yang-Baxter (YB) equations. The parameters $\eta, \alpha_\pm,\beta_\pm\theta_\pm$ are free and but are related to the various coupling constants, and interactions strengths in the problem at hand. Given these definitions the authours define the following monodromy and transfer matrices,
\begin{eqnarray}
\Theta_0(u)&=&K^+(u)R_{0N}(u+\theta_N)\dots R_{01}(u+\theta_1)K^-(u)R_{0N}(u-\theta_N)\dots R_{01}(u-\theta_1)\\
\tau(u)&=&\text{Tr}_0\,\Theta(u)
\end{eqnarray} 
following the Boundary inverse method we get $[\tau (u),\tau (v)]=0$ and thus the problem is tractable. Indeed they go on to construct the eigenvalues, $\Lambda (u)$ of $\tau (u)$ via an inhomogeneous T-Q relation. For even $N$ the result is 
\begin{eqnarray}\label{1}
\Lambda (u)=a(u)\frac{Q_1(u-\eta)}{Q_2(u)}+d(u)\frac{Q_2(u+\eta)}{Q_1(u)}+\frac{2\bar{c}\sinh{2u}\sinh{(2u+2\eta)}}{Q_1(u)Q_2(u)}A(u)A(-u-\eta)
\end{eqnarray}
Where the functions above are defined to be,
\begin{eqnarray}
A(u)&=&\prod^N_{l=1}\frac{\sinh{(u-\theta_l+\eta)}\sinh{(u+\theta_l+\eta)}}{\sinh^2{\eta}}\\
Q_1(u)&=&\prod^N_{j=1}\frac{\sinh{(u-\mu_j)}}{\sinh{\eta}},~~~~Q_2(u)=\prod^N_{j=1}\frac{\sinh{(u+\mu_j+\eta)}}{\sinh{\eta}}\\
a(u)&=&-4\frac{\sinh{(2u+2\eta)}}{\sinh{(2u+\eta)}}\sinh{(u-\alpha_-)}\cosh{(u-\beta_-)}\sinh{(u-\alpha_+)}\cosh{(u-\beta_+)}A(u)\\
d(u)&=&a(-u-\eta),~~\bar{c}=\cosh{\left[(N+1)\eta+\alpha_-+\alpha_++\beta_-+\beta_++2\sum^N_{j=1}\mu_j\right]}-\cosh{(\theta_--\theta_+)}
\end{eqnarray}
Here the parameters $\mu_j$ are the Bethe parameters and $\theta_l$ the inhomogeneities. From this T-Q relation one obtains the Bethe equations by demanding that the function has only simple poles whose residues vanish, which gives,
\begin{eqnarray}\nonumber
\frac{\bar{c}\sinh{(2\mu_j+\eta)}\sinh{(2\mu_j+2\eta)}}{2\sinh{(\mu_j+\alpha_-+\eta)}\cosh{(\mu_j+\beta_-+\eta)}\sinh{(\mu_j+\alpha_++\eta)}\cosh{(\mu_j+\beta_++\eta)}}\\
=\prod^N_{l=1}\frac{\sinh{(\mu_j+\mu_l+\eta)}\sinh{(\mu_j+\mu_l+2\eta)}}{\sinh{(\mu_j+\theta_l+\eta)}\sinh{(\mu_j-\theta_l+\eta)}}
\end{eqnarray}
Along with these we have so called selection rules $\mu_j\neq\mu_k$ and $\mu_j\neq-\mu_k-\eta$.

Now our problem is to diagonalise the operator
\[Z=S^{12}\dots S^{1N} S^{1}W^{1N}\dots W^{12}\]
in which \begin{equation}
S^{j}=\begin{pmatrix}
\alpha &&\beta\\
\beta && \alpha
\end{pmatrix},~~~~W^{ij}=\begin{pmatrix}
1&&0&&0&&0\\
0&&e^{i\phi}&&0&&0\\
0&&0&&e^{i\phi}&&0\\
0&&0&&0&&1
\end{pmatrix},~~\alpha=\frac{1-U^2/4}{1+U^2/4},~~\beta=\frac{-iU}{1+U^2/4},e^{i\phi}=\frac{1-ig}{1+ig}
\end{equation}
and $W^{ij}=P^{ij}$ is the permutation of the two spaces. In order to diagonalise this we introduce the R-matrix
\begin{eqnarray}
\mathcal{R}(u)=\begin{pmatrix}
1&&0&&0&&0\\
0&&\frac{\sinh{u}}{\sinh{(u+\eta)}}&&\frac{\sinh{\eta}}{\sinh{(u+\eta)}}&&0\\
0&&\frac{\sinh{\eta}}{\sinh{(u+\eta)}}&&\frac{\sinh{u}}{\sinh{(u+\eta)}}&&0\\
0&&0&&0&&1
\end{pmatrix}
\end{eqnarray}
which is related to both the S-matrices present in $Z$,
\begin{eqnarray}
\mathcal{R}_{ij}(0)=P^{ij},~~~~~\lim_{u\rightarrow\infty}\mathcal{R}_{ij}(u)|_{\eta=-i\phi}=W^{ij}
\end{eqnarray}
Thus we are lead to try diagonalise the transfer matrix, $t(u)$
\begin{eqnarray}
\Xi_0(u)&=&\mathcal{R}_{01}(u+\theta/2)\dots \mathcal{R}_{0N}(u+\theta/2)K^-(u)\mathcal{R}_{0N}(u-\theta/2)\dots \mathcal{R}_{0}(u-\theta/2)\\
t(u)&=&\text{Tr}_0\,\Xi(u)
\end{eqnarray}
Which is related to $Z$ by 
\begin{equation}
Z=\lim_{\theta\rightarrow\infty}t(\theta/2)
\end{equation}
provided the boundary matrix is chosen so that it goes to $S^0$ in the limit. We can see that $\Theta(u)$ and $\Xi(u)$ are similar in structure and indeed there is a simple mapping between them. Once we have this mapping then we can make the same replacements in \eqref{1}\eqref{2}and obtain the eigenvalues and bethe equations.

Firstly we should specify the boundary matrices. As there is no $K^+$ in $Z$ we should require that either $K^+=1$ or that it is $\propto 1$ when $u=\theta/2$ (or $B_0$ for twisted boundary conditions) and after $\lim_{\theta\rightarrow\infty}$. In addition $K^-$ should be proportional to $S^0$ after the same operations. Therefore we choose 
\begin{eqnarray}
K^{-}(u)&=&\frac{\beta}{\sinh{\theta}}\begin{pmatrix}
2i\cosh{(c+\theta/2)}\cosh{u}&& \sinh{2u}\\
\sinh{2u}&&2i\cosh{(c+\theta/2)}\cosh{u}
\end{pmatrix}\\
K^{+}(u)&=&\frac{e^{-\eta}}{\sinh{3\theta/2}}\begin{pmatrix}
2\left(\sinh{(-\theta)}\cosh{(i\Phi)}\cosh{(u+\eta)}\right. && -\sinh{(2u+2\eta)}\\
\,\left.-\cosh{(\theta)}\sinh{(i\Phi)}\sinh{(u+\eta)}\right)&& \\
&&2\left(\sinh{(-\theta)}\cosh{(i\Phi)}\cosh{(u+\eta)}\right.\\
-\sinh{(2u+2\eta)} &&\, \left.+\cosh{(\theta)}\sinh{(i\Phi)}\sinh{(u+\eta)}\right)
\end{pmatrix}
\end{eqnarray}
In both cases we have taken the liberty of including an overall constant factor. One can then check that 
\begin{eqnarray}
\lim_{\theta\rightarrow\infty}K^-(\theta/2)=\begin{pmatrix}
i\beta e^c&&\beta\\
\beta &&i\beta e^c
\end{pmatrix},~~~~\lim_{\theta\rightarrow\infty}K^+(\theta/2)=-\begin{pmatrix}
e^{i\Phi}&&0\\0&&e^{-i\Phi}
\end{pmatrix}
\end{eqnarray}
Which is what we want provided $e^c=\alpha/i\beta=(1-U^2/4)/U$. In terms of the parameters introduced previously, these are obtained by taking
\begin{equation}
\alpha_-=c+\theta/2+i\pi/2,~~~\alpha_+=-\theta,~~~\beta_-=0,~~~\beta_+=i\Phi,~~~\theta_\pm=0
\end{equation}
and including an overall factor of
\begin{equation}
\frac{-\beta e^{-\eta}}{\sinh{\theta}\sinh{3\theta/2}}
\end{equation}
Turning our attention to the $R$ matrices we see that they differ by an overall factor
\begin{equation}
\mathcal{R}(u)=\frac{\sinh\eta}{\sinh{(u+\eta)}}R(u)
\end{equation}
We are now able to relate $\Theta(u)$ and $\Xi (u)$. Specifically we want to go from $\Theta(u)$ to $\Xi(u)$. To achieve this relabel the spaces  so the orderings match, $N-m\rightarrow m+1$ and  take $\theta_k=\theta/2$ $\forall k$,
\begin{equation}
\Xi(u)=\frac{-\beta e^{-\eta}}{\sinh{\theta}\sinh{3\theta/2}}\prod^N_{j=1}\frac{\sinh\eta}{\sinh{(u-\theta/2+\eta)}}\frac{\sinh\eta}{\sinh{(u+\theta/2+\eta)}}\Theta(u)|_{\theta_k=\theta/2}
\end{equation}

We are interested in the eigenvalue at $u=\theta/2=\theta_j$. At this value of the spectral parameter the second and third terms in $\Lambda (u)$ vanish so we are merely interested in
\begin{equation}
\Lambda(\theta/2)=-4i\beta e^{-\eta}\frac{\sinh{(\theta+2\eta)}\cosh{(c)}\cosh{(\theta/2)}\cosh{(\theta/2+i\Phi)}}{\sinh{(\theta+\eta)}\sinh{\theta}}\prod^N_j\frac{\sinh{(\theta/2-\mu_j-\eta)}}{\sinh{(\theta/2+\mu_j+\eta)}}
\end{equation}

The Bethe equations are
\begin{eqnarray*}\nonumber
\frac{\left[\cosh{\left((N+1)\eta+c+i\pi/2+i\Phi-\theta/2+2\sum^N_{j=1}\mu_j\right)}-1\right]\sinh{(2\mu_j+\eta)}\sinh{(2\mu_j+2\eta)}}{2\sinh{(\mu_j+c+\theta/2+\eta)}\cosh{(\mu_j+\eta)}\cosh{(\mu_j+\eta+i\Phi)}\sinh{(\mu_j-\theta+\eta)}}\\
=\prod^N_{l=1}\frac{\sinh{(\mu_j+\mu_l+\eta)}\sinh{(\mu_j+\mu_l+2\eta)}}{\sinh{(\mu_j+\theta/2+\eta)}\sinh{(\mu_j-\theta/2+\eta)}}
\end{eqnarray*}

Up till now we have dealt with $N$ even however there also exists a solution for $N$ odd. This requires the use of $N+1$ Bethe parameters. The energy is still given by \eqref{Lambda} but with the sums running up to $(N+1)/2$. Additionally the Bethe equations are modified,

\begin{eqnarray*}\nonumber
\frac{\left[\cosh{\left((N+3) \eta+c+i\Phi-\theta/2+2\sum^{N+1}_{j=1}\mu_j\right)}-1\right]\sinh{(2\mu_j+\eta)}\sinh{(2\mu_j+2\eta)}\sinh{(\mu_j)}\sinh{(\mu_j+\eta)}}{2\sinh{(\mu_j+c+\theta/2+\eta)}\cosh{(\mu_j+\eta)}\cosh{(\mu_j+i\Phi+\eta)}\sinh{(\mu_j-\theta+\eta)}}\\
=\frac{1}{\sinh^{N}{(\mu_j+\theta/2+\eta)}\sinh^N{(\mu_j-\theta/2+\eta)}}\prod^{N+1}_{l=1}\sinh{(\mu_j+\mu_l+\eta)}\sinh{(\mu_j+\mu_l+2\eta)}
\end{eqnarray*}

\section*{Appendix B}
In this section we check that upon setting the impurity strength to zero that the solution reduces to the Luttinger Liquid. First we should describe the desired result. For a Luttinger liquid we can specify the number of left and right movers as they are conserved.  WLOG we can set the number of right movers to be $M$ and the number of left movers $N-M$. For one of the right movers to traverse the system on a ring of length it must scatter past the $N-M$ left movers and so it has a total phase shift $(N-M)i\phi$. Therefore the  right mover contribution to the energy is $-M(N-M)\phi/L$. Similarly a left mover has a total phase shift of $Mi\phi$ and therefore the left moving sector also contributes  $-M(N-M)\phi/L$. We should hope to find that the energy reduces to 
\begin{equation}
E=\dots-i\frac{2M(N-M)}{L}\eta
\end{equation}
Where $\eta=-i\phi$. In addition the degeneracy of this energy is $ {N \choose M} $. Now we look to our derived Bethe equations. We will only consider $N$ even but for $N$ odd the same argument applies. 
%\begin{eqnarray*}\nonumber
%\frac{\left[\cosh{\left((N+1)\eta+c-\theta/2+2\sum^N_{j=1}\mu_j\right)}-1\right]\sinh{(2\mu_j+\eta)}\sinh{(2\mu_j+2\eta)}}{2\sinh{(\mu_j+c+\theta/2+\eta)}\cosh^2{(\mu_j+\eta)}\sinh{(\mu_j-\theta+\eta)}}=\prod^N_{l=1}\frac{\sinh{(\mu_j+\mu_l+\eta)}\sinh{(\mu_j+\mu_l+2\eta)}}{\sinh{(\mu_j+\theta/2+\eta)}\sinh{(\mu_j-\theta/2+\eta)}}
%\end{eqnarray*}
%and that it stays the same order on both side if $\mu_j=\lambda_j+\varkappa/2$, $\mu_k=\nu_{k-N/2}-\varkappa$ for $k>N/2$. Which gave us
%\begin{eqnarray}\nonumber
% e^{2\lambda_j-2c}\sinh^N{(\lambda_j+\eta)}&=&e^{2\sum_k(2\lambda_k+\nu_k)}e^{3N\eta/2}\prod_{k=1}^{N/2}\sinh{(\lambda_j+\nu_k+\eta)}\sinh{(\lambda_j+\nu_k+2\eta)}
%\\\nonumber
%e^{-\nu_j-\eta-c}\frac{\sinh^N{(\nu_j+\eta)}}{2\sinh{(\nu _j+c+\eta)}}&=&\prod^{N/2}_{l=1}\sinh{(\nu_j+\lambda_l+\eta)}\sinh{(\nu_j+\lambda_l+2\eta)}e^{2\lambda_l+\eta}.
%\end{eqnarray}
%after taking the limit $\theta\rightarrow\infty$. 
Removing the impurity corresponds to $U=0$ or taking $c\rightarrow\infty$. We see that upon doing so the left hand side vanishes and we are forced to conclude that the Bethe roots form pairs $(\lambda_j,\nu_j)$, of two types,
\begin{equation}
(\lambda_j,-\lambda_j-\eta) ~~~\text{or}~~~~(\lambda_j,-\lambda_j-2\eta)
\end{equation}
In terms of of the original roots we have the condition that either $\mu_{j+N/2}=-\mu_j-\eta$ or $\mu_{j+N/2}=-\mu_j-2\eta$. However there are still $N/2$ free parameters $\mu_j$. To constrain these we need to use this pair structure in the T-Q relation. Let's say that there are $M$ pairs of roots such that $\mu_{j+N/2}=-\mu_j-\eta$ and that we reorder them so that these occur for $j=1\dots M$. We can then sub this back into our T-Q relation for the eigenvalue $\Lambda (u)$ and take the limit $c\rightarrow \infty$. Our new T-Q relation is
\begin{eqnarray}\nonumber
\Lambda (u)&=&\frac{-2e^{\theta/2-u-\eta}}{\sinh{\theta}\sinh{3\theta/2}}\frac{\sinh(2u+2\eta)}{\sinh{(2u+\eta)}}\sinh{(u+\theta)}\cosh{u}\cosh{(u-i\Phi)}\prod^M\frac{\sinh{(u-\mu_j-\eta)}}{\sinh{(u+\mu_j+\eta)}}\frac{\sinh{(u+\mu_j)}}{\sinh{(u-\mu_j)}}\\\nonumber
&+&\frac{2e^{\theta/2+u}}{\sinh{\theta}\sinh{3\theta/2}}\frac{\sinh 2u}{\sinh{(2u+\eta)}}\sinh{(u+\eta-\theta)}\frac{\sinh^N{(u-\theta/2)}\sinh^N{(u+\theta/2)}}{\sinh^N{(u-\theta/2+\eta)}\sinh^N{(u+\theta/2+\eta)}}\\
&&~~~~~~~~~~~~~~~~~~~~~~~~~~~~~~~~\times\cosh{(u+\eta)}\cosh{(u+i\Phi+\eta)}\prod^M\frac{\sinh{(u-\mu_j+\eta)}}{\sinh{(u+\mu_j+2\eta)}}\frac{\sinh{(u+\mu_j+\eta)}}{\sinh{(u-\mu_j)}}
\end{eqnarray}
There are two things to note about this expression the first is that the $N-M$ pair of roots of the second type have cancelled out and do not contribute and also the inhomogeneous term has also vanished. As before we are only interested in taking the eigenvalues at $u=\theta/2$ and then in the limit $\theta/2\rightarrow\infty$. With this value of the spectral parameter the second term also vanishes,
\begin{equation}
\Lambda (\theta/2)=\frac{2e^{-\eta}}{\sinh{\theta}}\frac{\sinh(\theta+2\eta)}{\sinh{\theta+\eta}}\cosh{(\theta/2)}\cosh{(\theta/2-i\Phi)}\prod^M\frac{\sinh{(\mu_j+\eta-\theta/2)}}{\sinh{(\mu_j-\theta/2)}}\frac{\sinh{(\mu_j+\theta/2)}}{\sinh{(\mu_j+\theta/2+\eta)}}
\end{equation}
If we shift $\mu_j=\lambda_j+\theta/2-\eta/2$ and take $\theta\rightarrow\infty$  get the momenta of the system
\begin{equation}
e^{-ikL}=e^{-M\eta-i\Phi}\prod^M\frac{\sinh{(\lambda_j+\eta/2)}}{\sinh{(\lambda_j-\eta/2)}}
\end{equation}
from which we get that the energy is given by 
\begin{equation}
E=\frac{2\pi}{L}\sum^N_kn_k+i\frac{N}{L}\sum^M\log\frac{\sinh{(\lambda_j+\eta/2)}}{\sinh{(\lambda_j-\eta/2)}}-i\frac{MN}{L}\eta+\frac{N}{L}\Phi
\end{equation}
To evaluate this explicitly we need to use the Bethe equations from our new T-Q relation. As before we demand that $\Lambda$ has only simple poles and that the residues vanish. The simple pole restriction gives us the selection rule $\mu_j\neq\mu_k$ and $\mu_j\neq-\mu_k-\eta$. The vanishing residues then results in the Bethe equations
\begin{eqnarray}\nonumber
0&=&\frac{-2e^{\theta/2-\mu_j-\eta}}{\sinh{\theta}\sinh{3\theta/2}}\frac{\sinh(2\mu_j+2\eta)}{\sinh{(2\mu_j+\eta)}}\sinh{(\mu_j+\theta)}\cosh{(\mu_j)}\cosh{(\mu_j-i\Phi)}\prod^M\frac{\sinh{(\mu_j-\mu_k-\eta)}}{\sinh{(\mu_j+\mu_k+\eta)}}\sinh{(\mu_j+\mu_k)}\\\nonumber
&+&\frac{2e^{\theta/2+\mu_j}}{\sinh{\theta}\sinh{3\theta/2}}\frac{\sinh 2\mu_j}{\sinh{(2\mu_j+\eta)}}\sinh{(\mu_j+\eta-\theta)}\frac{\sinh^N{(\mu_j-\theta/2)}\sinh^N{(\mu_j+\theta/2)}}{\sinh^N{(\mu_j-\theta/2+\eta)}\sinh^N{(\mu_j+\theta/2+\eta)}}\\
&&~~~~~~~~~~~~~~~~~~~~~~~~~~~\times\cosh{(\mu_j+\eta)}\cosh{(\mu_j+i\Phi+\eta)}\prod^M\frac{\sinh{(\mu_j-\mu_k+\eta)}}{\sinh{(\mu_j+\mu_k+2\eta)}}\sinh{(\mu_j+\mu_k+\eta)}
\end{eqnarray}
Performing the necessary algebra give us
\begin{eqnarray}\nonumber
e^{-2\mu_j-\eta}\frac{\sinh(2\mu_j+2\eta)\sinh{(\mu_j+\theta)}\cosh{(\mu_j)}\cosh{(\mu_j-i\Phi)}}{\sinh 2\mu_j\sinh{(\mu_j+\eta-\theta)}\cosh{(\mu_j+\eta)}\cosh{(\mu_j+i\Phi+\eta)}}\frac{\sinh^N{(\mu_j-\theta/2+\eta)}\sinh^N{(\mu_j+\theta/2+\eta)}}{\sinh^N{(\mu_j-\theta/2)}\sinh^N{(\mu_j+\theta/2)}}\\
=\prod^M\frac{\sinh{(\mu_j-\mu_k+\eta)}\sinh{(\mu_j+\mu_k+\eta)}}{\sinh{(\mu_j-\mu_k-\eta)}\sinh{(\mu_j+\mu_k)}}
\end{eqnarray}
We should make the same change of variables as before. Here do it in two steps for clarity. First let $\mu_j=\lambda_j+\theta/2$,
\begin{eqnarray}\nonumber
\frac{\sinh(2\lambda_j+\theta+2\eta)\sinh{(\lambda_j+3\theta/2)}\cosh{(\lambda_j+\theta/2)}\cosh{(\lambda_j+\theta/2-i\Phi)}}{\sinh{(2\lambda_j+\theta)}\sinh{(\mu_j+\eta-\theta)}\cosh{(\lambda_j+\theta/2+\eta)}\cosh{(\lambda_j+\theta/2+i\Phi+\eta)}}\frac{\sinh^N{(\lambda_j+\eta)}\sinh^N{(\lambda_j+\theta+\eta)}}{\sinh^N{\lambda_j}\sinh^N{(\lambda_j+\theta)}}\\
=e^{-2\lambda_j-\theta-\eta}\prod^M\frac{\sinh{(\lambda_j-\lambda_k+\eta)}\sinh{(\lambda_j+\lambda_k+\theta+\eta)}}{\sinh{(\lambda_j-\lambda_k-\eta)}\sinh{(\lambda_j+\lambda_k+\theta)}}
\end{eqnarray}
Now we can take the limit and shift $\lambda_j$ by $-\eta/2$ and get \eqref{llBae}
\begin{equation}
\frac{\sinh^N{(\lambda_j+\eta/2)}}{\sinh^N{(\lambda_j-\eta/2)}}e^{N\eta-2i\Phi}=-e^{2M\eta}\prod^M\frac{\sinh{(\lambda_j-\lambda_k+\eta)}}{\sinh{(\lambda_j-\lambda_k-\eta)}}.
\end{equation}
Taking the $\log$ of these we obtain
\begin{equation}\label{2}
N\log{\frac{\sinh{(\lambda_j+\eta/2)}}{\sinh{(\lambda_j-\eta/2)}}}=-(N-2M)\eta+2i\Phi+\sum_k^M\log{\frac{\sinh{(\lambda_j-\lambda_k+\eta)}}{\sinh{(\lambda_j-\lambda_k-\eta)}}}+2\pi i I_j
\end{equation}
where $I_j$ is a half integer. Using this in our energy equation and the fact that we have a double sum over the antisymmetric function $\log{\frac{\sinh{(\lambda_j-\lambda_k+\eta)}}{\sinh{(\lambda_j-\lambda_k-\eta)}}}$ we get \eqref{lle}.
\end{widetext}
\end{document}